# Internal resonances and dynamic responses in equivalent mechanical model of partially liquid-filled vessel

M. Farid and O. V. Gendelman[*]

Faculty of Mechanical Engineering, Technion – Israel Institute of Technology

*contacting author, ovgend@tx.technion.ac.il

## Abstract

The paper treats oscillations of a liquid in partially filled vessel under horizontal harmonic ground excitation. Such excitation may lead to hydraulic impacts. The liquid sloshing mass is modeled by equivalent pendulum, which can impact the vessel walls. We use parameters of the equivalent pendulum for well-explored case of cylindrical vessels. The hydraulic impacts are modeled by high-power potential function. Conditions for internal resonances are presented. A non-resonant behavior and dynamic response related to 3:1 internal resonance are explored. When the excitation amplitude exceeds a critical value, the system exhibits multiple steady state solutions. Quasi-periodic solutions appear in relatively narrow range of parameters. Numerical continuation links between resonant regimes found asymptotically for small excitation amplitude, and high-amplitude responses with intensive impacts.

## 1. Introduction

Vessels filled with liquid are used in many fields of engineering, including nuclear [1], vehicle [2,3] and aerospace industries [4], for storage of chemicals, gasoline, water, and various hazardous liquids [5]. External excitations may cause well-known dynamical effect of liquid sloshing. This effect can take place in liquid cargo on highways, cruises or in stationary vessels exposed to earthquakes. Dynamic loads related to the liquid sloshing may have direct and rather strong hazardous effect on the vessel stability and robustness.

So far, detailed analytical explorations are limited to small-amplitude sloshing in rectangular and cylindrical vessels. While being most interesting and potentially hazardous, high-amplitude liquid sloshing in cylindrical tanks still lacks complete analytic description. The reason is that the liquid in the tank is continuous system with infinite number of degrees of freedom, and its boundary conditions on the free surface are nonlinear and time-dependent. Nevertheless, loads created by high-amplitude liquid sloshing are so crucial for designing the containing vessel, its supports and payload limitations [6], that a number of approximate phenomenological models were suggested in order to get at least qualitative insight into this phenomenon.

In one of the most well-known phenomenological models, the sloshing dynamics in a partially-filled liquid tank with total mass $M$ is modeled by a pendulum with mass $m$, length of $l$, and a rotational coordinate $\theta$ with respect to the vessel center-line. In this simplified model, three dynamic regimes can take place, as shown in Pilipchuk and Ibrahim[7] and presented in Fig. 1:

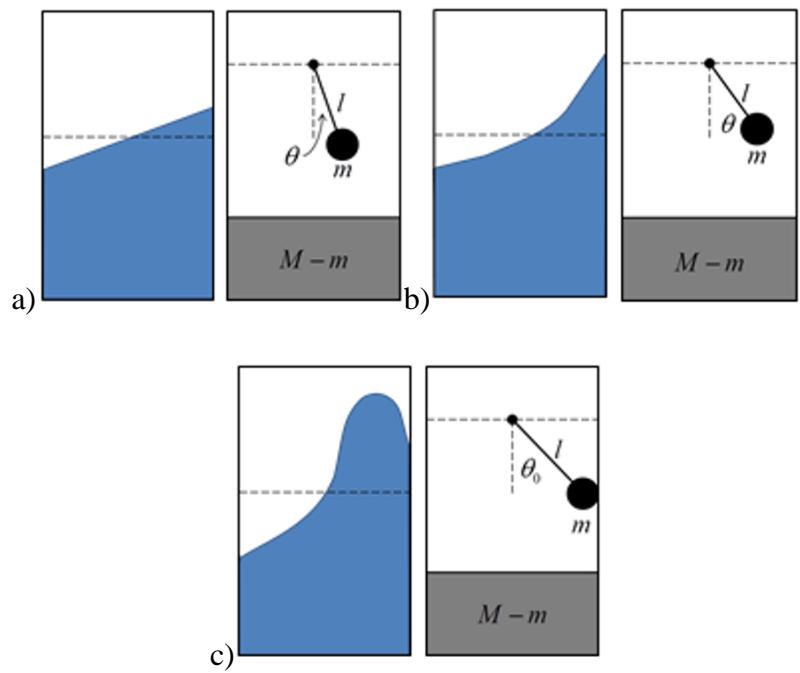

**Fig. 1.** Regimes of liquid free-surface motion and their equivalent mechanical models.

(a) The liquid free surface performs small oscillations around its trivial stable equilibrium and remains planar. This regime can be successfully described by a pendulum performing small oscillations.
(b) Relatively large oscillations in which the liquid free surface does not remain planar. This motion is described by a differential equation with weak nonlinearity, and can be treated by perturbation methods [7–9]. In this regime the "equivalent" spherical pendulum is considered to perform moderate oscillations, so that approximation $\sin\theta \approx \theta - \frac{\theta^3}{3!}$ is reasonable, and the nonlinearity can be treated as weak.

(c) The free liquid surface is urged into a strongly nonlinear motion, related to liquid sloshing impacts with the tank walls. This regime can be described with the help of a pendulum, which impacts the tank walls.

High-amplitude sloshing can cause hydraulic jumps. In this case (corresponding to Fig. 1c) major hydraulic impacts can act on the vessel structure walls [10]. Despite obvious practical interest, methods for evaluation of the impact in this case are not well developed, and rely primarily on the data of direct experiments and simplified phenomenological models [11]. Hydraulic jumps and wave collisions with vessel walls give rise to strongly nonlinear behavior in the system, since collisions with rigid or elastic masses cause rapid velocity changes. Hence, even if the sloshing appears due to simple harmonic excitations, the response may be neither harmonic nor periodic. Authors of paper [12] suggest application of methods developed for analysis of vibro-impact motion. In this work another approach is applied. We use high order smooth potential function to model the interaction between the free-surface wave and the vessel walls, following Pilipchuk[13,14], Pilipchuk and Ibrahim [7] and Gendelman[15]. Smooth polynomial potential function was used by Pilipchuk and Ibrahim[16] to model small amplitude free liquid oscillations in elevated rigid container. The vibro-impact problem of a pendulum oscillating in a rigid container was treated in previous works of Buzhinskii and Stolbetsov [17] and by Shaw and Shaw[18].

Modeling of free-surface oscillations in rectangular tanks with the help of the equivalent pendulum was suggested by Graham in 1951 [19]. Equivalent moment of inertia of a liquid in cylindrical containers has been estimated numerically by Partom ([20] and [21]) and verified experimentally by Werner and Coldwell [22]. Parameters of the equivalent pendulum, which corresponds to the first asymmetric sloshing mode of cylindrical and rectangular tanks, were studied by Dodge [23] and Abramson [11]. The non-linear interaction of the sloshing liquid with elastic tank in conditions of parametric excitation was studied by Ibrahim[24], Ibrahim and Barr[25,26], Ibrahim,

Gau and Soundararajan[27], and El-Sayad, Hanna and Ibrahim [28]. Observations and experiments show that in the vicinity of internal resonances violent response might take place. Using a continuous liquid model, similar interactions resulting from horizontal and combined ground excitation were studied by Ibrahim and Li[29], and for parametric excitation by Soundarajan and Ibrahim[30]. Both harmonic parametric excitation and saw-tooth approximation for the horizontal external excitation were applied by Pilipchuk and Ibrahim [7].

Dynamics of the liquid sloshing in partially filled vessels depends on the depth of the liquid. For shallow free-surface, waves and hydraulic jumps reveal themselves in a vicinity of resonance and apply high loads on vessel walls [10]. For high free-surface, hydraulic loads may be applied on the vessel top.

Horizontal ground excitation may lead to variety of liquid dynamic regimes, such as hydraulic impacts and fluid swirling around the tank vertical axis. However, in this work we focus on the former regime due to its significant contribution to the stresses formation in the tank structure. Thus, we consider relatively simple equivalent mechanical model with two degrees of freedom that mimics the dynamic regime of interest. In order to derive the parameters of the linear liquid sloshing one should refer one of the models proposed by Abramson [11] for rectangular or cylindrical tank, and to calculate the relevant parameters accordingly. We decided to deal with the special case of cylindrical tanks and to explore the response of equivalent mechanical model of partially liquid-filled vessel in conditions of 1:1 primary resonance and 3:1 internal resonance. It seems to be a substantial extension of previous results, since the internal resonance between the sloshing liquid and the tank can lead to a notable increase of the response amplitude. Conditions for the internal resonances are checked for various sets of system parameters. Dynamics of the system in the case of high-amplitude base excitation is also examined with the help of numerical continuation approach.

This paper is organized as follows: in Section2 we describe the model and develop its governing equations of motion. Section3 contains analytic exploration of the model equilibria and conditions for the internal resonances. Then, asymptotic multiple – scales analysis is applied for investigation of the forced response. Numerical results of the forced system are presented in Section 4 for the non-resonant case and for the 3:1 internal resonance. Numeric continuation relates these regimes to high-amplitude sloshing.

## 2. Description of the model and governing equations

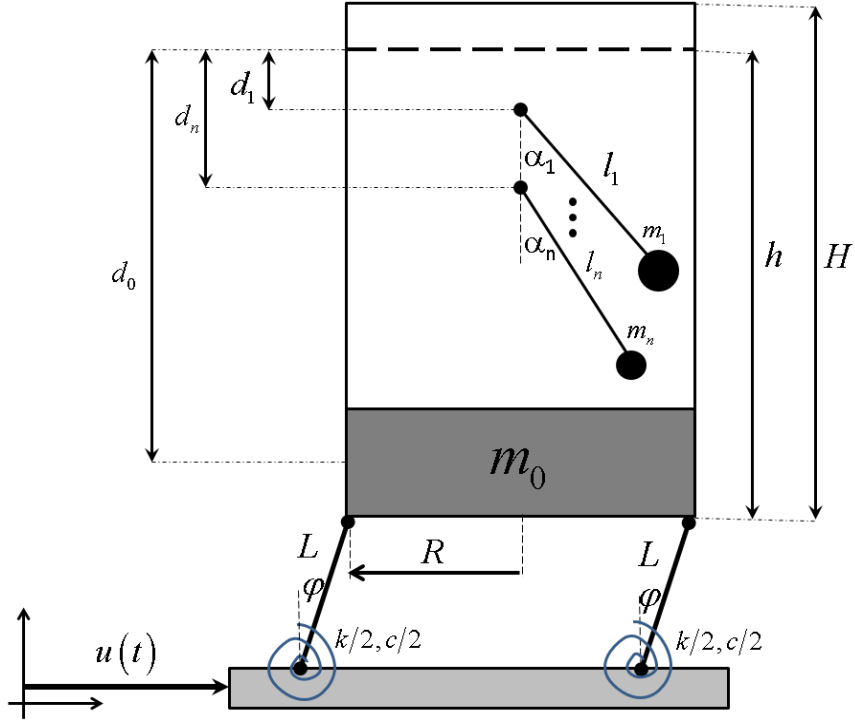

**Fig. 2.** Scheme of multiple sloshing modes in the cylindrical tank. The modes are represented by series of pendula interacting with the walls.

In the following Section, following Ibrahim [31], we introduce the equivalent mechanical model for liquid sloshing in cylindrical tank, shown in **Fig. 2**. Cylindrical container supported by massless rods with length L is considered. The rods are restrained by torsional springs with overall stiffness coefficient $k$ and linear viscous damping coefficient $c$. These parameters mimic both elasticity and natural damping of the tank shell. The undisturbed free surface of the liquid has height $h$ with respect to the bottom of the vessel, and is presented by a bold dashed line. It is supposed that the base of the tank performs horizontal motion with displacement $u(t)$; $M$ is the total mass of the container-liquid system. The sloshing model involves an infinite series of pendula, and the $n^{th}$ pendulum represents the $n^{th}$ sloshing mode. Each pendulum has effective mass $m_n$, length $l_n$ and angle with respect to the vessel center-line $\alpha_n$. A rigid mass $m_0$ represents the liquid portion that does not perform sloshing, and acts as a rigid body at the bottom of the vessel. Mass preservation yields:

$$m_F = m_0 + \sum_{n=1}^{\infty} m_n \tag{1}$$

For liquid with density $\rho$, the total liquid mass is $m_F = \pi \rho R^2 h$. As shown by Dodge [23], the mass of each modal pendulum decreases rapidly with increasing mode number. Consequently, it is reasonable to take into account only the first mode in the mechanical equivalent model, as long as the excitation frequency is far from the natural frequencies of the higher modes.

For cylindrical vessel with radius $R$, liquid depth of $h$ and total mass of the liquid $m_F$, the length, mass and location with respect to the undisturbed free-surface of the fundamental mode pendulum is given by the following expressions, as documented by Abramson [11]:

$$l = \frac{R}{1.84} \coth(1.84 h/R), \; m = m_F \frac{R}{2.2h} \tanh(1.84 h/R), \; d_1 = \frac{R}{3.68} \csc h(3.68 h/R) \tag{2}$$

The non-sloshing liquid portion mass $m_0$ and location below the undisturbed liquid free-surface $d_0$ are as follows:

$$d_0 = \frac{m_F}{m_0}\left(\frac{h}{2} - \frac{R^2}{2h}\right) - (d_1 + l)\frac{m}{m_0}, \; m_0 = m_F - m \tag{3}$$

The angle between the pendulum and the tank axis is further denoted by $\theta$. Following Dodge [23] we assume that the pendulum is always perpendicular to the liquid free-surface and the total liquid mass $m_F = m + m_0$. The latter assumption is not completely accurate due to the presence of additional sloshing modes. We also assume that the pendulum hits the tank walls when angle θ achieves values $\pm\theta_0$. The angle between the tank rods and the vertical axis is denoted by $\varphi$.

Interaction between the pendulum and the tank walls is described by the following strongly nonlinear power-form forces with high exponents potential function [7,13–15]:

$$U_{impact} = \frac{b\theta_0}{2n}\left(\frac{\theta}{\theta_0}\right)^{2n} \quad (3)$$

where $b$ is a positive coefficient which is determined experimentally and $n \gg 1$ is a positive integer.

The corresponding force field represents the impact force applied by the liquid on the tank as a function of $\theta$, and is given by the following expression:

$$F_{impact} = \frac{d}{d\theta}(U_{impact}) = b\left(\frac{\theta}{\theta_0}\right)^{2n-1} \quad (4)$$

This interaction force between the tank walls and the pendulum for different values of $n$ is exemplified in Fig. 3 as follows:

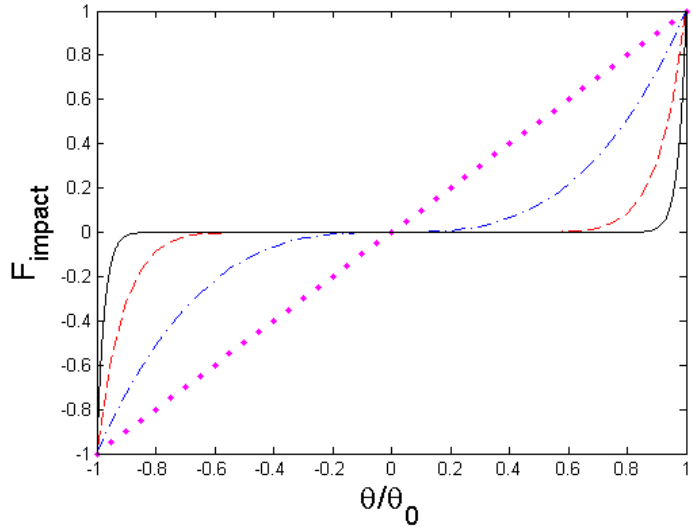

**Fig. 3.** The interaction force between the tank walls and the pendulum vs. the pendulum angle for n=1(dotted pink), n=3(dashed-dotted blue), n=6(dashed red) and n=10(solid black).

Impact of the sloshing liquid at tank walls corresponds to the limit $|\theta| \to \theta_0$. Note that for $n \to \infty$ the potential function corresponds to the square potential well. As parameter $n$ reaches large, but finite values, the potential function remains finite at $\theta = \pm\theta_0$ and further. One can say that this property of the model potential mimics the fact that the sloshing wave and tank walls are deformable rather than rigid. We further follow Ibrahim et al. [31] and take $n=6$.

The liquid-structure interaction involves energy dissipation due to wave breaking and fluid viscosity. Following Pilipchuk and Ibrahim [7], the following approximate dissipation force is used in the model:

$$F_d = d\left(\frac{\theta}{\theta_0}\right)^{2p} \dot{\theta} \quad (5)$$

In this formula $d$ is a constant coefficient which is determined experimentally, $p$ is a positive integer (generally, $p \neq n$), and $\dot{\theta}$ is the angular velocity of the pendulum. Following El-Sayad, Hanna and Ibrahim [28], we take $p=2$. Similar dissipation functions can be used for simulation of the vibro-impact systems, as mentioned in [32].

The equations of motion governing the liquid-filled vessel system are formulated with the help of Lagrange formalism. Kinetic and potential energies of the system are expressed as:

$$T = \frac{M}{2}L^2\dot{\varphi}^2 + \frac{m}{2}l^2\dot{\theta}^2 + mlL\dot{\theta}\dot{\varphi}\cos(\theta+\varphi) + \dot{u}\left(\frac{M}{2} + ML\dot{\varphi}\cos\varphi + ml\dot{\theta}\cos\theta\right)$$
$$V = \frac{k}{2}\varphi^2 + MgL\cos\varphi + mg(H - l\cos\theta) + \frac{b\theta_0}{2n\theta_0^{2n}}\theta^{2n} \quad (6)$$

Hence, the Lagrangian is given by the following expression:

$$L = T - V = \frac{M}{2}L^2\dot{\varphi}^2 + \frac{m}{2}l^2\dot{\theta}^2 + mlL\dot{\theta}\dot{\varphi}\cos(\theta+\varphi) + \dot{u}\left(\frac{M}{2} + ML\dot{\varphi}\cos\varphi + ml\dot{\theta}\cos\theta\right)$$
$$- \frac{k}{2}\varphi^2 - MgL\cos\varphi - mg(H - l\cos\theta) - \frac{b\theta_0}{2n\theta_0^{2n}}\theta^{2n} \quad (7)$$

The virtual work applied by the dissipation forces is expressed as:

$$dW_d = -c\dot{\varphi}d\varphi - d\left(\frac{\theta}{\theta_0}\right)^{2p}\dot{\theta}d\theta \quad (8)$$

The following non-dimensional parameters governing the system properties are defined:

$$s_1 = \frac{l}{L}, \mu_1 = \frac{m}{M}, \kappa = \frac{k}{\omega^2 L^2 M}$$
$$\beta = \frac{b}{\omega^2 m l^2}, X = \frac{u}{L}, \gamma = \frac{c}{\omega L^2 M}, \delta = \frac{d}{\omega m l^2} \qquad (9)$$

where $\omega^2 = g/l$ is the natural frequency of the pendulum. Note that $\kappa$ represents the ratio between the vessel and pendulum natural frequencies, and $\mu_1$ is the pendulum and vessel mass ratio. Thus, by using the extended Hamilton's principle, the non-dimensional equations of motion are written as follows:

$$\ddot{x}_1 + a_1 \cos(\theta_0(x_1 + x_2))\ddot{x}_2 = a_2 \sin(\theta_0(x_1 + x_2))\dot{x}_2^2 - a_3 \sin(\theta_0 x_1)$$
$$- a_4 x_1^{2n-1} - a_5 x_1^{2p} \dot{x}_1 - a_6 \cos(\theta_0 x_1) \ddot{X}$$

$$\ddot{x}_2 + b_1 \cos(\theta_0(x_1 + x_2))\ddot{x}_1 = b_2 \sin(\theta_0(x_1 + x_2))\dot{x}_1^2 - b_3 x_2 + b_4 \sin(\theta_0 x_2)$$
$$- b_5 \dot{x}_2 - b_6 \ddot{X} \cos(\theta_0 x_2) \qquad (10)$$

where $x_1 = \theta/\theta_0$, $x_2 = \varphi/\theta_0$, the dots represent differentiation with respect to the normalized time $\tau = \omega t$, and the normalized coefficients are as follows:

$$a_1 = \frac{1}{s_1}, a_2 = \frac{\theta_0}{s_1}, a_3 = \frac{1}{\theta_0}, a_4 = \frac{\beta}{\theta_0}, a_5 = \delta, a_6 = \frac{1}{\theta_0 s_1}$$
$$b_1 = \mu_1 s_1, b_2 = \mu_1 s_1 \theta_0, b_3 = \kappa, b_4 = \frac{s_1}{\theta_0}, b_5 = \gamma, b_6 = \frac{1}{\theta_0} \qquad (11)$$

Equations (10) can be cast in the following matrix form:

$$\mathbf{M}\ddot{\mathbf{Q}} = \mathbf{F} \qquad (12)$$

Here $\mathbf{Q} = \begin{bmatrix} x_1 & x_2 \end{bmatrix}^T$ is the state vector, $\mathbf{M}$ is the inertia matrix, and $\mathbf{F}$ is the forcing vector, a function of the state variables and their first order derivatives.

The inertia matrix $\mathbf{M}$ of the system is written as:

$$\mathbf{M} = \begin{pmatrix} 1 & a_1 \cos(\theta_0(x_1 + x_2)) \\ b_1 \cos(\theta_0(x_1 + x_2)) & 1 \end{pmatrix} \qquad (13)$$

Note that $\det(\mathbf{M}) = 1 - \mu_1 \cos^2(\theta_0(x_1 + x_2)) > 0$ for every value of $0 < \mu_1 < 1$ and therefore matrix $\mathbf{M}$ is nonsingular. The forcing vector $\mathbf{F}$ is presented as follows:

$$\mathbf{F} = \begin{pmatrix} a_2 \sin(\theta_0(x_1 + x_2))\dot{x}_2^2 - a_3 \sin(\theta_0 x_1) - a_4 x_1^{2n-1} - a_5 x_1^{2p} \dot{x}_1 - a_6 \cos(\theta_0 x_1)\ddot{X} \\ b_2 \sin(\theta_0(x_1 + x_2))\dot{x}_1^2 - b_3 x_2 + b_4 \sin(\theta_0 x_2) - b_5 \dot{x}_2 - b_8 \ddot{X} \cos(\theta_0 x_2) \end{pmatrix} \quad (14)$$

To perform numerical analysis, we diagonalize equation(12) by pre-multiplying it by matrix $\mathbf{M}^{-1}$ to obtain the following inertially uncoupled equations of motion:

$$\ddot{\mathbf{Q}} = \mathbf{M}^{-1}\mathbf{F} \quad (15)$$

## 3. Analytic treatment

### 3.1. States of equilibrium

The equilibrium points of system(10) are obtained by eliminating the derivatives with respect to the normalized time $\tau$:

$$\begin{aligned} a_3 \sin(\theta_0 x_1) + a_4 x_1^{2n-1} &= 0 \\ -b_3 x_2 + b_4 \sin(\theta_0 x_2) &= 0 \end{aligned} \quad (16)$$

The first equation of system(16) yields:

$$\sin\theta = -\hat{\beta}\theta^{2n-1} \quad (17)$$

where $\hat{\beta} = \dfrac{\beta}{\theta_0^{2n-1}}$. The equilibrium points with respect to $\theta$ are computed from equation (17) and shown graphically in Fig. 4 or several values of $\hat{\beta}$:

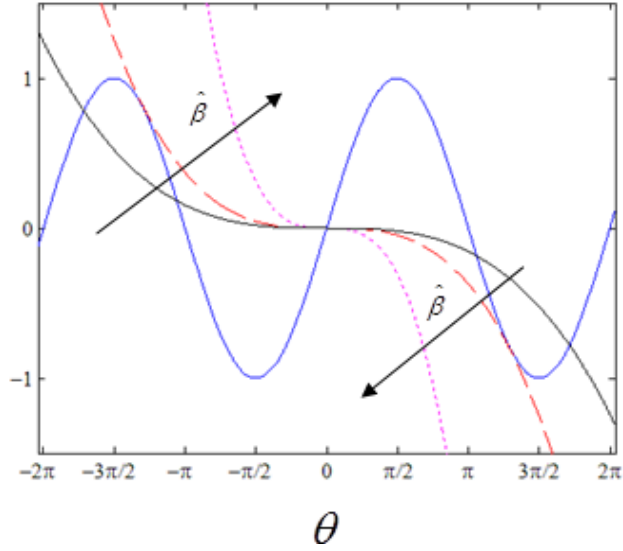

**Fig. 4.** Equilibrium points of $\theta$ respresnted by intersection between $\sin\theta$ (solid blue) and $-\hat{\beta}\theta^{2n-1}$ for $n=2$, $\hat{\beta}=0.005$ (solid black), $\hat{\beta}=0.012$ (dashed red) and $\hat{\beta}=0.08$ (dotted magenta).

As one can see in **Fig. 4**, for critical value of parameter $\hat{\beta}=\hat{\beta}_b$ bifurcation takes place, i.e. quantitative change in the equilibrium points of coordinate $\theta$. As shown in **Fig. 4**, $\hat{\beta}_b = 0.012$. The dotted line is produced for $\hat{\beta} > \hat{\beta}_b$, the solid line for $\hat{\beta} < \hat{\beta}_b$ and the dashed line for $\hat{\beta} = \hat{\beta}_b$. In the case of $\hat{\beta} > \hat{\beta}_b$, there is a single intersection between both curves, which corresponds to a single trivial equilibrium point, in which the pendulum is in vertical position. For $\hat{\beta} < \hat{\beta}_b$, there are five equilibrium points- the trivial one, and two symmetric couples of additional points. The case of $\hat{\beta}=\hat{\beta}_b$ corresponds to a transition between the single equilibrium point to multiple equilibrium points. At this limiting case, the system formally has three distinct equilibrium points. Since $\theta_0 < \pi/2$ for every vessel design, the additional equilibrium points are not physical.

We now analyze the equilibrium solution for $\varphi$ using the second equation of system(16):

$$\sin\varphi = \hat{\kappa}\varphi \qquad (18)$$

where $\hat{\kappa} = \dfrac{\kappa}{s_1}$ is proportional to the dominance of structure stiffness with respect to the overall system inertia, described by κ. The equilibrium points derived from equation (18) are shown graphically in **Fig. 5** with respect to coordinate $\varphi$.

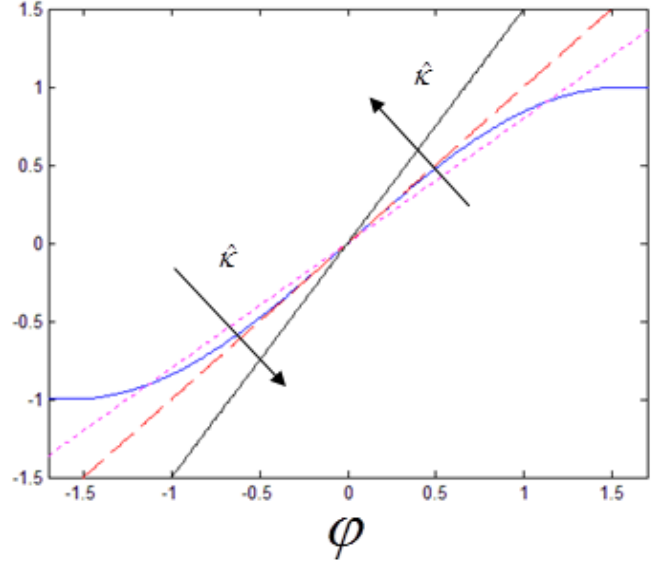

**Fig. 5.** Equilibrium points for variable $\varphi$ respresnted by intersection between $\sin\varphi$ (solid blue) and $\hat{\kappa}\varphi$ for $\hat{\kappa}=1.3$ (solid black), $\hat{\kappa}=1$ (dashed red) and $\hat{\kappa}=0.9$ (dotted magenta).

The solid line corresponds to $\hat{\kappa}>1$, and the dotted line for $\hat{\kappa}<1$. In the case of $\hat{\kappa}>1$, there is a single trivial equilibrium point, in which the vessel rods are vertical. In the case of $\hat{\kappa}<1$, there are three equilibrium points- the trivial one, and two symmetric additional points. In this case the vessel rods are inclined due to a balance between the torsional stiffness and gravity force. The dashed line, which corresponds to $\hat{\kappa}=1$, indicates a bifurcation point where a transition between the single equilibrium point to three equilibrium points takes place. The stability of these equilibrium points is analyzed in the following section.

### 3.1.1. Stability of trivial equilibrium

The Jacobian corresponding to the trivial equilibrium $\theta=\varphi=0$ may be written as follows:

$$\mathbf{J}_{trivial} = \begin{pmatrix} 0 & 1 & 0 & 0 \\ -\dfrac{1}{1-\mu_1} & 0 & \dfrac{\kappa - s_1}{s_1(1-\mu_1)} & \dfrac{\gamma}{s_1(1-\mu_1)} \\ 0 & 0 & 0 & 1 \\ \dfrac{\mu_1 s_1}{1-\mu_1} & 0 & -\dfrac{\kappa - s_1}{1-\mu_1} & -\dfrac{\gamma}{1-\mu_1} \end{pmatrix} \quad (19)$$

The corresponding characteristic polynomial is:

$$\lambda^4 + \frac{\gamma}{1-\mu_1}\lambda^3 + \frac{\kappa - s_1 + 1}{1-\mu_1}\lambda^2 + \frac{\gamma}{1-\mu_1}\lambda + \frac{\kappa - s_1}{1-\mu_1} = 0 \quad (20)$$

Note that for every vessel, the sloshing portion mass is smaller than the total vessel mass, and therefore $\mu_1 < 1$. Hence, according to Routh's criterion, the trivial equilibrium is stable for $\kappa > s_1$, or $\hat{\kappa} > 1$.

### 3.1.2. Nontrivial Equilibrium

The Jacobian corresponding to the nontrivial equilibrium $\theta = 0$, $\varphi \neq 0$ has the form:

$$\mathbf{J}_{NonTrivial} = \begin{pmatrix} 0 & 1 & 0 & 0 \\ J_{21} & 0 & J_{23} & J_{24} \\ 0 & 0 & 0 & 1 \\ J_{41} & 0 & J_{43} & J_{44} \end{pmatrix} \quad (21)$$

where coefficients $J_{ij}$ are expressed as follows:

$$J_{21} = \frac{s_1\theta_0 + \sin(\theta_0 x_3)\theta_0^2 \kappa x_3 - \sin^2(\theta_0 x_3)\theta_0 s_1}{s_1\theta_0(\mu_1 \cos^2(\theta_0 x_3) - 1)}$$

$$+ 2\mu_1 \cos(2\theta_0)\sin(2\theta_0)\frac{s_1 \cos(2\theta_0)\sin(2\theta_0) - 2\kappa\theta_0 \cos(2\theta_0)}{s_1(\mu_1 \cos^2(\theta_0 x_3) - 1)^2}$$

$$J_{23} = \frac{\sin(\theta_0 x_3)\theta_0^2 \kappa x_3 - \theta_0 \kappa \cos(\theta_0 x_3) - \sin^2(\theta_0 x_3)\theta_0 s_1 + s_1\theta_0 \cos^2(\theta_0 x_3)}{s_1\theta_0(\mu_1 \cos^2(\theta_0 x_3) - 1)}$$

$$+ 2\mu_1 \cos(\theta_0 x_3)\sin(\theta_0 x_3)\frac{s_1 \cos(\theta_0 x_3)\sin(\theta_0 x_3) - \kappa\theta_0 x_3 \cos(\theta_0 x_3)}{s_1(\mu_1 \cos^2(\theta_0 x_3) - 1)^2}$$

$$J_{41} = -\frac{\mu_1 s_1 \cos(\theta_0 x_3)}{\mu_1 \cos^2(\theta_0 x_3) - 1} + 2\mu_1 \cos(\theta_0 x_3)\sin(\theta_0 x_3)\frac{\kappa\theta_0 x_3 - s_1 \sin(\theta_0 x_3)}{(\mu_1 \cos^2(\theta_0 x_3) - 1)^2}$$

$$J_{43} = \frac{\kappa\theta_0 - s_1\theta_0 \cos(\theta_0 x_3)}{(\mu_1 \cos^2(\theta_0 x_3) - 1)\theta_0} + 2\mu_1 \cos(\theta_0 x_3)\sin(\theta_0 x_3)\frac{\kappa\theta_0 x_3 - s_1 \sin(\theta_0 x_3)}{(\mu_1 \cos^2(\theta_0 x_3) - 1)^2}$$

$$J_{24} = \frac{-\gamma \cos(\theta_0 x_3)}{s_1\theta_0(\mu_1 \cos^2(\theta_0 x_3) - 1)}$$

$$J_{44} = \frac{\gamma}{\mu_1 \cos^2(\theta_0 x_3) - 1}$$

(22)

The corresponding characteristic polynomial looks as:

$$\lambda^4 - J_{44}\lambda^3 - (J_{21} + J_{43})\lambda^2 + (J_{21}J_{44} - J_{24}J_{41})\lambda + (J_{21}J_{43} - J_{23}J_{41}) \tag{23}$$

The nontrivial equilibrium points are stable according to Routh criterion for $\hat{\kappa} < 1$. The analytical predictions of the equilibrium points and their stability are shown in **Fig. 6**.

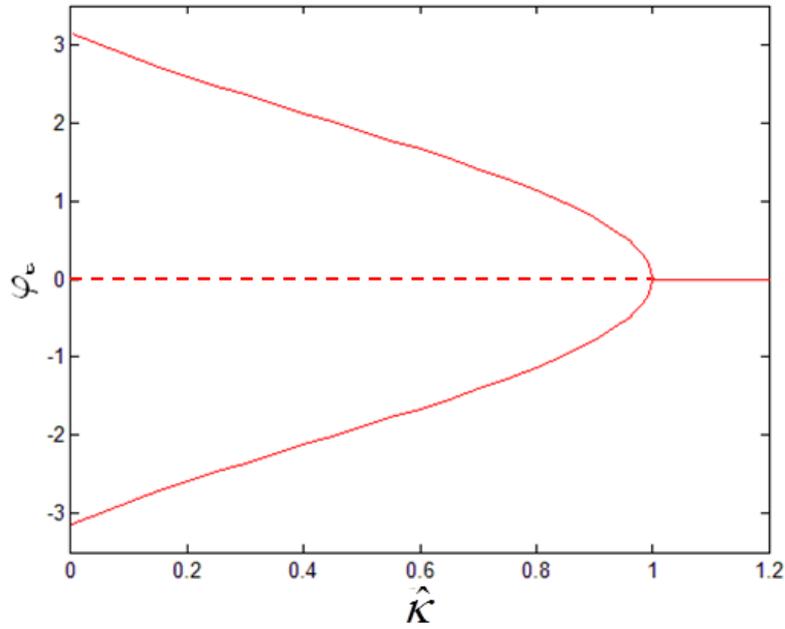

**Fig. 6.** Stable (solid) and unstable (dashed) equilibrium solutions of φ versus $\hat{\kappa}$.

We learn from Fig. 6 that for larger values of vessel rods stiffness, i.e. $\hat{\kappa} > 1$, the trivial equilibrium point is stable. However, for $\hat{\kappa} < 1$ the trivial equilibrium point becomes unstable, and a new pair of stable nontrivial equilibrium points shows up due to supercritical pitchfork bifurcation for $\hat{\kappa} = 1$.

Physically, this loss of stability may be identified with overthrow of the vessel. Then, we assume that all values of $\hat{\kappa}$ are larger than unity, and the trivial equilibrium point is stable.

### 3.2. Conditions of internal resonances

Natural frequencies of the system are calculated by considering the unforced ($A=0$) and undamped ($\gamma = 0$) case and performing linearization in the vicinity of trivial equilibrium. The linearized equations of motion are given by Jacobian of System (19), according to the following equation:

$$\begin{pmatrix} \dot{x}_1 \\ \ddot{x}_1 \\ \dot{x}_2 \\ \ddot{x}_2 \end{pmatrix} = \mathbf{J}_{trivial} \begin{pmatrix} x_1 \\ \dot{x}_1 \\ x_2 \\ \dot{x}_2 \end{pmatrix} \quad (24)$$

where $\mathbf{J}_{trivial}$ is given in equation (19). In order to obtain linearized equations of motion, we use only the second derivative expressions in equation (26) and get the following system of equations:

$$\begin{pmatrix} \ddot{x}_1 \\ \ddot{x}_2 \end{pmatrix} + \mathbf{K} \begin{pmatrix} x_1 \\ x_2 \end{pmatrix} = \begin{pmatrix} 0 \\ 0 \end{pmatrix} \quad (25)$$

Here $\mathbf{K}$ is the linear stiffness matrix and is calculated as follows:

$$\mathbf{K} = \begin{pmatrix} k_{11} & -k_{12} \\ -k_{21} & k_{22} \end{pmatrix} \quad (26)$$

Expressions for coefficients $k_{ij}$ are given in appendix A. The system natural linear frequencies $\omega_1, \omega_2$ fulfill the following equation:

$$\left| \mathbf{K} - \omega_{1,2}^2 \mathbf{I} \right| = 0 \quad (27)$$

where $\mathbf{I}$ is the identity matrix. The natural frequencies of the system are calculated explicitly by the following expressions:

$$\omega_1 = \sqrt{\hat{\alpha} - \sqrt{\hat{\alpha}^2 - \hat{\beta}}}, \quad \omega_2 = \sqrt{\hat{\alpha} + \sqrt{\hat{\alpha}^2 - \hat{\beta}}} \quad (28)$$

with $\hat{\alpha} = \dfrac{1 - s_1 + \kappa}{2(1 - \mu_1)}$ and $\hat{\beta} = \dfrac{\kappa - s_1}{1 - \mu_1}$.

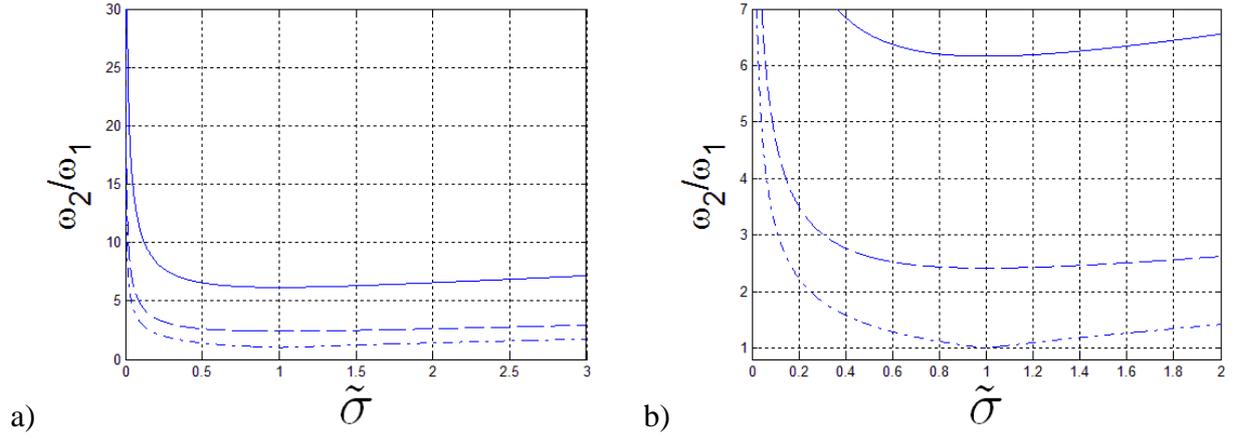

**Fig. 7.** a) Natural friquencies ratio vs $\tilde{\sigma}$ for $\tilde{\mu}=0.1$ (solid), $\tilde{\mu}=0.5$ (dashed) and $\tilde{\mu}=1$ (dashed-dotted), b) Zoom-in for $0<\tilde{\sigma}<2$.

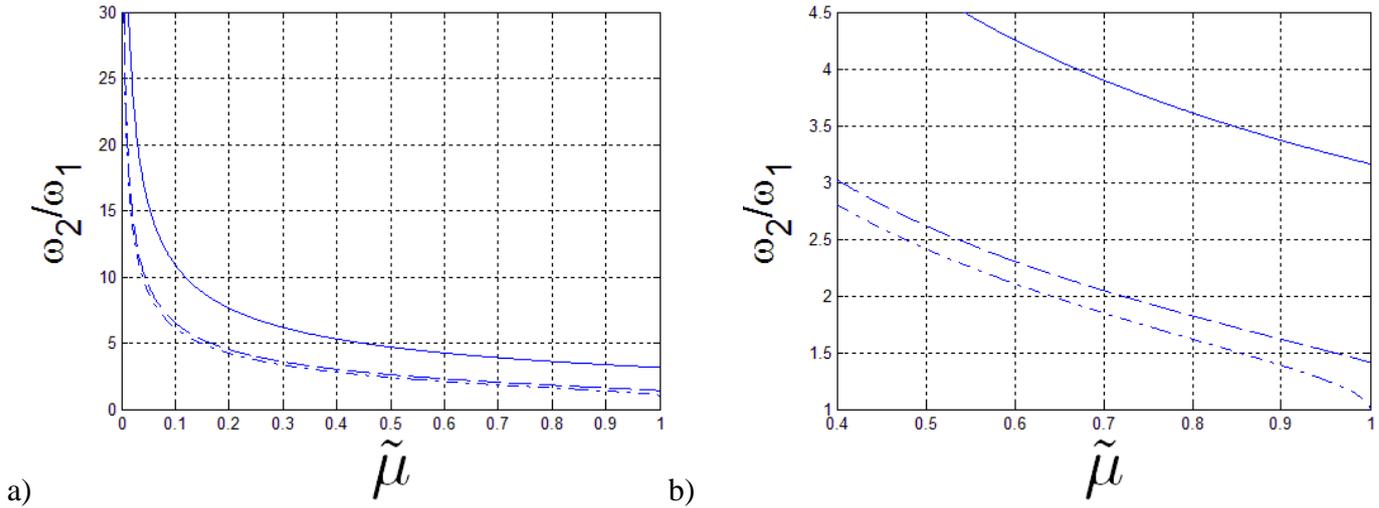

**Fig. 8.** a) Natural friquencies ratio vs. $\tilde{\mu}$ for $\tilde{\sigma}=0.1$ (solid), $\tilde{\sigma}=0.5$ (dashed) and $\tilde{\sigma}=1$ (dashed-dotted), b) Zoom-in for $0.4<\tilde{\mu}<1$.

The ratio between the system natural frequencies versus several system parameters is presented in Fig. 7 and Fig. 8. Fig. 7 shows the natural frequency ratio $\omega_2/\omega_1$ versus the non-dimensional parameter $\tilde{\sigma}=\kappa-s_1$ for distinct values of parameter $\tilde{\mu}=1-\mu_1$ while Fig. 8 shows this ratio versus parameter $\tilde{\mu}$ for distinct values of $\tilde{\sigma}$. Since the sloshing mass is significantly smaller than the total mass of the full tank, parameter $\mu_1$ and parameter $\tilde{\mu}$ are much smaller than unity. For feasible values of $\tilde{\mu}$ the 1:1 internal resonance cannot occur. However, other internal resonances can take place. For example, one can observe from Fig. 7 that for $\tilde{\mu}=0.5, \tilde{\sigma}\approx 0.305$ and 3:1 internal resonance is possible. This interesting particular case is analyzed in the next Section.

### 3.3. Asymptotic analysis

Let us consider now the case of base excitation in horizontal direction. If this excitation is harmonic, namely $u(t) = A\sin(\Omega t)$, the equations of motion are expressed in the following form:

$$\ddot{x}_1 + a_1 \cos(\theta_0(x_1+x_2))\ddot{x}_2 = a_2 \sin(\theta_0(x_1+x_2))\dot{x}_2^2 - a_3 \sin(\theta_0 x_1)$$
$$-a_4 x_1^{2n-1} - a_5 x_1^{2p}\dot{x}_1 + \rho^2 X a_6 \cos(\theta_0 x_1)$$

$$\ddot{x}_2 + b_1 \cos(\theta_0(x_1+x_2))\ddot{x}_1 = b_2 \sin(\theta_0(x_1+x_2))\dot{x}_1^2 - b_3 x_2 + b_4 \sin(\theta_0 x_2)$$
$$-b_5 \dot{x}_2 + \rho^2 X b_6 \cos(\theta_0 x_2)$$

(29)

In this system $\rho = \Omega/\omega$ and $X=A/L$. System of equations (29) can be written in the matrix form:

$$\mathbf{M}\begin{pmatrix}\ddot{x}_1 \\ \ddot{x}_2\end{pmatrix} = \begin{pmatrix}\tilde{f}_1 \\ \tilde{f}_2\end{pmatrix} \tag{30}$$

where the matrix of inertia $\mathbf{M}$ is given by equation (13), and the terms $\tilde{f}_1$ and $\tilde{f}_2$ are expressed as follows:

$$\tilde{f}_1 = a_2 \sin(\theta_0(x_1+x_2))\dot{x}_2^2 - a_3 \sin(\theta_0 x_1) - a_4 x_1^{2n-1} - a_5 x_1^{2p}\dot{x}_1 + \rho^2 A a_6 \sin(\rho\tau)\cos(\theta_0 x_1)$$
$$\tilde{f}_2 = b_2 \sin(\theta_0(x_1+x_2))\dot{x}_1^2 - b_3 x_2 + b_4 \sin(\theta_0 x_2) - b_5 \dot{x}_2 + \rho^2 A b_6 \sin(\rho\tau)\cos(\theta_0 x_2)$$

(31)

As mentioned in section 2, the matrix of inertia $\mathbf{M}$ is non-singular. We further multiply Equation (30) by matrix $\mathbf{M}^{-1}$ to get the inertially uncoupled equations of motion. Expansion of the resulting system of equations into Taylor series up to cubic terms around the trivial stable equilibrium state yields the following system of equations:

$$\begin{pmatrix}\ddot{x}_1 \\ \ddot{x}_2\end{pmatrix} + \mathbf{K}\begin{pmatrix}x_1 \\ x_2\end{pmatrix} = \begin{pmatrix}g_1 \\ g_2\end{pmatrix} \tag{32}$$

where the linear stiffness matrix $\mathbf{K}$ is defined in equation (26). Note that since the linearized equations are valid only for small oscillations, the contribution of hydraulic impacts is neglected at this stage. The expressions of $g_1$ and $g_2$ are as follows:

$$g_1 = \hat{\gamma}_{11}\dot{x}_2 + \hat{\alpha}_{11}x_1^3 + \hat{\alpha}_{12}x_1^2 x_2 - \hat{\gamma}_{12}x_1^2 \dot{x}_2 - \hat{\gamma}_{13}\dot{x}_1^2 x_1 + \hat{\alpha}_{13}x_1 x_2^2 - \hat{\gamma}_{14}x_1 x_2 \dot{x}_2$$
$$+ \hat{\gamma}_{15}x_1\dot{x}_2^2 - \hat{\gamma}_{16}x_2\dot{x}_1^2 + \hat{\alpha}_{14}x_2^3 - \hat{\gamma}_{17}x_2^2\dot{x}_2 + \hat{\gamma}_{18}x_2\dot{x}_2^2 + \rho^2\left(\hat{F}_{12}x_1 x_2 + \hat{F}_{13}x_2^2\right)A\sin(\rho\tau)$$
(33)

$$g_2 = -\hat{\gamma}_{21}\dot{x}_2 - \hat{\alpha}_{21}x_1^3 + \hat{\alpha}_{22}x_1^2 x_2 + \hat{\gamma}_{22}x_1^2\dot{x}_2 + \hat{\gamma}_{23}\dot{x}_1^2 x_1 + \hat{\alpha}_{23}x_1 x_2^2 + \hat{\gamma}_{24}x_1 x_2 \dot{x}_2$$
$$- \hat{\gamma}_{25}x_1\dot{x}_2^2 + \hat{\gamma}_{26}x_2\dot{x}_1^2 + \hat{\alpha}_{24}x_2^3 + \hat{\gamma}_{27}x_2^2\dot{x}_2 - \hat{\gamma}_{28}x_2\dot{x}_2^2 + \rho^2\left(\hat{F}_{21} - \hat{F}_{22}x_1 x_2 - \hat{F}_{23}x_2^2\right)A\sin(\rho\tau)$$

Expressions for coefficients $\hat{\gamma}_{ij}, \hat{\alpha}_{ij}, \hat{F}_{ij}$ are presented in Appendix A.

The system is reduced to modal coordinates through transformation: $\mathbf{x} = \mathbf{P}\mathbf{u}$ where $\mathbf{P}$ is the transformation matrix, which reduces matrix $\mathbf{P}^{-1}\mathbf{KP}$ to Jordan canonical form:

$$\mathbf{P} = \begin{pmatrix} -k_{12} & -k_{12} \\ \omega_1^2 - k_{11} & \omega_2^2 - k_{11} \end{pmatrix}, \quad \mathbf{P}^{-1} = \frac{1}{\omega_2^2 - \omega_1^2}\begin{pmatrix} -\dfrac{\omega_2^2 - k_{11}}{k_{12}} & -1 \\ \dfrac{\omega_1^2 - k_{11}}{k_{12}} & 1 \end{pmatrix}$$
(34)

After this transformation, system of equations (32) is reduced to the following linearly uncoupled system of equations:

$$\ddot{u}_1 + \omega_1^2 u_1 = h_1(u_1, u_2, \dot{u}_1, \dot{u}_2)$$
$$\ddot{u}_2 + \omega_2^2 u_2 = h_2(u_1, u_2, \dot{u}_1, \dot{u}_2)$$
(35)

In these equations, $\omega_1, \omega_2$ are equal to those in equation (28) and $h_i$ are given by the following expressions:

$$\omega_1^2 = \frac{1}{2}(k_{11} + k_{22}) - \frac{1}{2}\sqrt{(k_{11} - k_{22})^2 + 4k_{12}k_{21}}; \quad \omega_2^2 = \frac{1}{2}(k_{11} + k_{22}) + \frac{1}{2}\sqrt{(k_{11} - k_{22})^2 + 4k_{12}k_{21}}$$

$$h_1 = -\gamma_{11}\dot{u}_1 - \gamma_{12}\dot{u}_2 - \alpha_{11}u_1^3 - \alpha_{12}u_2^3 - \alpha_{13}u_1^2 u_2 - \alpha_{14}u_1 u_2^2 - \gamma_{13}u_1^2 \dot{u}_2$$
$$- \gamma_{14}u_2^2\dot{u}_2 - \gamma_{15}u_1^2\dot{u}_1 - \gamma_{16}u_2^2\dot{u}_1 - \gamma_{17}u_1\dot{u}_1^2 - \gamma_{18}u_1\dot{u}_1\dot{u}_2 - \gamma_{19}u_1\dot{u}_2^2$$
$$- \gamma_{110}u_2\dot{u}_1\dot{u}_2 - \gamma_{111}u_2\dot{u}_1^2 - \gamma_{112}u_1 u_2\dot{u}_1 - \gamma_{113}u_2\dot{u}_2^2 - \gamma_{114}u_1 u_2\dot{u}_2$$
$$+ \rho^2\left(-F_{11} + F_{12}u_1 u_2 + F_{13}u_2^2 + F_{14}u_1^2\right)A\sin(\rho\tau)$$
(36)

$$h_2 = -\gamma_{21}\dot{u}_1 - \gamma_{22}\dot{u}_2 - \alpha_{21}u_1^3 - \alpha_{22}u_2^3 - \alpha_{23}u_1^2 u_2 - \alpha_{24}u_1 u_2^2 - \gamma_{23}u_1^2\dot{u}_2$$
$$- \gamma_{24}u_2^2\dot{u}_2 - \gamma_{25}u_1^2\dot{u}_1 - \gamma_{26}u_2^2\dot{u}_1 - \gamma_{27}u_1\dot{u}_1^2 - \gamma_{28}u_1\dot{u}_1\dot{u}_2 - \gamma_{29}u_1\dot{u}_2^2$$
$$- \gamma_{210}u_2\dot{u}_1\dot{u}_2 - \gamma_{211}u_2\dot{u}_1^2 - \gamma_{212}u_1 u_2\dot{u}_1 - \gamma_{213}u_2\dot{u}_2^2 - \gamma_{214}u_1 u_2\dot{u}_2$$
$$+ \rho^2\left(F_{11} + F_{22}u_1 u_2 + F_{23}u_2^2 + F_{24}u_1^2\right)A\sin(\rho\tau)$$

Expressions for coefficients $\gamma_{ij}, \alpha_{ij}, F_{ij}$ are presented in Appendix A. Following Nayfeh and Mook [33], we look for a solution of the following form:

$$u_1(\tau) = \varepsilon u_{11}(T_0, T_2) + \varepsilon^3 u_{13}(T_0, T_2)$$
$$u_2(\tau) = \varepsilon u_{21}(T_0, T_2) + \varepsilon^3 u_{23}(T_0, T_2)$$
(37)

where $\varepsilon$ is a small parameter and $T_i = \varepsilon^i \tau$.

The time derivatives are expressed as: $\dfrac{d}{d\tau} = D_0 + \varepsilon^2 D_2 + \ldots \quad \dfrac{d^2}{d\tau^2} = D_0^2 + 2\varepsilon^2 D_0 D_2 + \ldots$

$D_i = \dfrac{\partial}{\partial T_i}$ and $D_i^2 = \dfrac{\partial^2}{\partial T_i^2}$.

The following rescaling of the damping coefficient and external excitation amplitude is adopted:

$$\gamma = \varepsilon^2 \bar{\gamma}, A = \varepsilon^3 \bar{A} \tag{38}$$

Thus, following appendix A, the respective damping coefficients are rescaled as follows:

$$\gamma_{11} = \varepsilon^2 \bar{\gamma}_{11}, \gamma_{12} = \varepsilon^2 \bar{\gamma}_{12}, \gamma_{13} = \varepsilon^2 \bar{\gamma}_{13}, \gamma_{14} = \varepsilon^2 \bar{\gamma}_{14}, \gamma_{15} = \varepsilon^2 \bar{\gamma}_{15}, \gamma_{16} = \varepsilon^2 \bar{\gamma}_{16}, \gamma_{112} = \varepsilon^2 \bar{\gamma}_{112}, \gamma_{114} = \varepsilon^2 \bar{\gamma}_{114}$$
$$\gamma_{21} = \varepsilon^2 \bar{\gamma}_{21}, \gamma_{22} = \varepsilon^2 \bar{\gamma}_{22}, \gamma_{23} = \varepsilon^2 \bar{\gamma}_{23}, \gamma_{24} = \varepsilon^2 \bar{\gamma}_{24}, \gamma_{25} = \varepsilon^2 \bar{\gamma}_{25}, \gamma_{26} = \varepsilon^2 \bar{\gamma}_{26}, \gamma_{212} = \varepsilon^2 \bar{\gamma}_{212}, \gamma_{214} = \varepsilon^2 \bar{\gamma}_{214}$$
(39)

Substituting equations (37)-(39) into equation (35), and collecting equal-order equations with respect to $\varepsilon$, one obtains sets of equations for successive powers of the small parameter:

$O(\varepsilon)$:

$$\ddot{u}_{11} + \omega_1^2 u_{11} = 0$$
$$\ddot{u}_{21} + \omega_2^2 u_{21} = 0$$
(40)

$O(\varepsilon^3)$:

$$\ddot{u}_{13} + \omega_1^2 u_{13} = -2D_0 D_2 u_{11} - \bar{\gamma}_{11}(D_0 u_{11}) - \bar{\gamma}_{12}(D_0 u_{21}) - \gamma_{17} u_{11}(D_0 u_{11})^2$$
$$- \gamma_{18} u_{11}(D_0 u_{11})(D_0 u_{21}) - \gamma_{19} u_{11}(D_0 u_{21})^2 - \gamma_{110} u_{21}(D_0 u_{11})(D_0 u_{21})$$
$$- \gamma_{111} u_{21}(D_0 u_{11})^2 - \gamma_{113} u_{21}(D_0 u_{21})^2$$
$$- \alpha_{11} u_{11}^3 - \alpha_{12} u_{21}^3 - \alpha_{13} u_{11}^2 u_{21} - \alpha_{14} u_{11} u_{21}^2$$
$$+ \frac{i}{2} F_{11} \rho^2 \bar{A} e^{i\rho T_0} - \frac{i}{2} F_{11} \rho^2 \bar{A} e^{-i\rho T_0}$$

(41)

$$\ddot{u}_{23} + \omega_2^2 u_{23} = -2D_0 D_2 u_{21} - \bar{\gamma}_{21}(D_0 u_{11}) - \bar{\gamma}_{22}(D_0 u_{21}) - \gamma_{27} u_{11}(D_0 u_{11})^2$$
$$- \gamma_{28} u_{11}(D_0 u_{11})(D_0 u_{21}) - \gamma_{29} u_{11}(D_0 u_{21})^2 - \gamma_{210} u_{21}(D_0 u_{11})(D_0 u_{21})$$
$$- \gamma_{211} u_{21}(D_0 u_{11})^2 - \gamma_{213} u_{21}(D_0 u_{21})^2$$
$$- \alpha_{21} u_{11}^3 - \alpha_{22} u_{21}^3 - \alpha_{23} u_{11}^2 u_{21} - \alpha_{24} u_{11} u_{21}^2$$
$$- \frac{i}{2} F_{11} \rho^2 \bar{A} e^{i\rho T_0} + \frac{i}{2} F_{11} \rho^2 \bar{A} e^{-i\rho T_0}$$

System of equations (40) yields:

$$u_{11} = A_1(T_2) e^{i\omega_1 T_0} + \bar{A}_1(T_2) e^{-i\omega_1 T_0}$$
$$u_{21} = A_2(T_2) e^{i\omega_2 T_0} + \bar{A}_2(T_2) e^{-i\omega_2 T_0}$$

(42)

Equation (42) is substituted into equation (41), and then the secular terms are eliminated. We distinguish between two different cases-the non-resonant case, where $\omega_2$ is far from $3\omega_1$, and internal resonance of 3:1 case, where $\omega_2 \approx 3\omega_1$.

### 3.3.1. Non-Resonant case

In that case, the system is far from internal resonance of 3:1 and all other internal resonances are considered to be weak. The secular terms in equations (41) are eliminated to get the solvability conditions. However, since we look for solutions providing a primary resonance of 1:1 with respect to $\omega_1$, the following detuning parameter is considered:

$$\varepsilon^2 \sigma = \rho - \omega_1$$

(43)

Then the solvability conditions are written as follows:

$$-2i\omega_1 D_2 A_1 - i\omega_1 \bar{\gamma}_{11} A_1 - \left(\gamma_{17}\omega_1^2 + 3\alpha_{11}\right) A_1^2 \bar{A}_1 - 2\left(\gamma_{19}\omega_2^2 + \alpha_{14}\right) A_1 A_2 \bar{A}_2$$
$$+ \frac{i}{2}\rho^2 \bar{A} F_{11} e^{i\sigma T_2} = 0 \tag{44}$$
$$-2i\omega_2 D_2 A_2 - i\omega_2 \bar{\gamma}_{22} A_2 - \left(\gamma_{213}\omega_2^2 + 3\alpha_{22}\right) A_2^2 \bar{A}_2 - 2\left(\gamma_{211}\omega_1^2 + \alpha_{23}\right) A_1 \bar{A}_1 A_2 = 0$$

The following traditional polar transformations are applied:

$$A_1(T_2) = \frac{1}{2} a_1(T_2) e^{i\theta_1(T_2)}, \bar{A}_1(T_2) = \frac{1}{2} a_1(T_2) e^{-i\theta_1(T_2)}$$
$$A_2(T_2) = \frac{1}{2} a_2(T_2) e^{i\theta_2(T_2)}, \bar{A}_2(T_2) = \frac{1}{2} a_2(T_2) e^{-i\theta_2(T_2)} \tag{45}$$

This polar transformation is introduced into equation (44) and real and imaginary parts are separated to yield the following slowly varying evolution equations:

$$D_2 a_1 = -\frac{1}{2}\bar{\gamma}_{11} a_1 + \frac{\rho^2 \bar{A} F_{11}}{2\omega_1} \cos\phi$$
$$a_1 D_2 \phi = \sigma a_1 - \frac{1}{8\omega_1}\left(c_{11} a_1^2 + 2 c_{12} a_2^2\right) a_1 - \frac{\rho^2 \bar{A} F_{11}}{2\omega_1} \sin\phi \tag{46}$$
$$D_2 a_2 = -\frac{1}{2}\bar{\gamma}_{22} a_2$$
$$a_2 D_2 \theta_2 = \frac{1}{8\omega_2}\left(c_{21} a_2^2 + 2 c_{22} a_1^2\right) a_2$$

where the expression for $\phi$ is given by:

$$\phi = \sigma T_2 - \theta_1 \tag{47}$$

The third equation of system (46) can be solved analytically:

$$a_2 = a_{2,0} e^{-\frac{\bar{\gamma}_{22}}{2} T_2} \tag{48}$$

where $a_{2,0} = a_2(T_2 = 0)$. Steady-state solutions are searched by eliminating the time derivatives from Equation (46). In this case, as one can see from Equation (48), $a_2$ goes to zero. We isolate $\sin\phi$ and $\cos\phi$, squaring, adding and equating to unity to yield the following bi-cubic polynomial in $a_1$, which can be solved analytically:

$$\hat{F}^2 = a_1^2 \left( \hat{\gamma}_1^2 + \left( \sigma - \frac{c_{11}}{8\omega_1} a_1^2 \right)^2 \right) \tag{49}$$

where $\hat{F} = \frac{\rho^2 \bar{A} F_{11}}{2\omega_1}$ and $\hat{\gamma}_1 = \frac{\gamma_1}{2}$. After substituting back to the original parameters, equation(49) is transformed to the following bi-cubic polynomial in $|u_1| = \varepsilon a_1$:

$$\frac{\rho^4 A^2 F_{11}^2}{4\omega_1^2} = \varepsilon a_1 \left( \frac{\gamma_{11}^2}{4} + \left( \rho - \omega_1 - \frac{c_{11}}{8\omega_1} (\varepsilon a_1)^2 \right)^2 \right) \tag{50}$$

The frequency responses corresponding obtained from equation(50) are shown in section 4.1.

### 3.3.2. 3:1 Internal Resonance

In this case we look for solutions corresponding to the internal resonance of 3:1 and primary resonance 1:1 with respect to $\omega_1$, and define the following detuning parameters:

$$\begin{aligned} \varepsilon^2 \sigma_1 &= \omega_2 - 3\omega_1 \\ \varepsilon^2 \sigma_2 &= \rho - \omega_1 \end{aligned} \tag{51}$$

Hence, the solvability conditions are as follows:

$$\begin{aligned} &-2i\omega_1 D_2 A_1 - i\omega_1 \bar{\gamma}_{11} A_1 - \left( \gamma_{17} \omega_1^2 + 3\alpha_{11} \right) A_1^2 \bar{A}_1 - 2\left( \gamma_{19} \omega_2^2 + \alpha_{14} \right) A_1 A_2 \bar{A}_2 \\ &\quad + \left( \gamma_{111} \omega_1^2 - \gamma_{18} \omega_1 \omega_2 - \alpha_{13} \right) \bar{A}_1^2 A_2 e^{i\sigma_1 T_2} + \frac{i}{2} \rho^2 \bar{A} F_{11} e^{i\sigma_2 T_2} = 0 \\ &-2i\omega_2 D_2 A_2 - i\omega_2 \bar{\gamma}_{22} A_2 - \left( \gamma_{213} \omega_2^2 + 3\alpha_{22} \right) A_2^2 \bar{A}_2 - 2\left( \gamma_{211} \omega_1^2 + \alpha_{23} \right) A_1 \bar{A}_1 A_2 \\ &\quad + \left( \gamma_{27} \omega_1^2 - \alpha_{21} \right) A_1^3 e^{-i\sigma_1 T_2} = 0 \end{aligned} \tag{52}$$

Substituting the polar transformations shown in equation (45) into equation (52), and separating real and imaginary parts, one obtains the following slow-flow equations:

$$D_2 a_1 = -\frac{1}{2}\bar{\gamma}_{11} a_1 + \frac{c_{13}}{8\omega_1} a_1^2 a_2 \sin\phi_1 + \frac{F_{11}}{2\omega_1}\rho^2 \bar{A}\cos\phi_2$$

$$D_2\phi_1 = \sigma_1 + \left(\frac{c_{21}}{8\omega_2} - \frac{3c_{12}}{4\omega_1}\right)a_2^2 + \left(\frac{c_{22}}{4\omega_2} - \frac{3c_{11}}{8\omega_1}\right)a_1^2 - \frac{c_{23}}{8\omega_2}\frac{a_1^3}{a_2}\cos\phi_1 + \frac{3c_{13}}{8\omega_1}a_1 a_2 \cos\phi_1 - \frac{3F_{11}}{2\omega_1 a_1}\rho^2 \bar{A}\sin\phi_2 \quad (53)$$

$$D_2 a_2 = -\frac{1}{2}\bar{\gamma}_{22} a_2 - \frac{c_{23}}{8\omega_2} a_1^3 \sin\phi_1$$

$$D_2\phi_2 = \sigma_2 - \frac{c_{11}}{8\omega_1}a_1^2 - \frac{c_{12}}{4\omega_1}a_2^2 + \frac{c_{13}}{8\omega_1}a_1 a_2 \cos\phi_1 - \frac{F_{11}}{2\omega_1 a_1}\rho^2 \bar{A}\sin\phi_2$$

Coefficients $c_{ij}$ are defined in Appendix A, and the expressions for $\phi_1$ and $\phi_2$ are as follows:

$$\phi_1 = \sigma_1 T_2 + \theta_2 - 3\theta_1, \quad \phi_2 = \sigma_2 T_2 - \theta_1 \quad (54)$$

Steady state solutions are given by eliminating by eliminating the time derivatives of equation (53). After simple algebraic manipulation, equation (53) yields two polynomial system of $a_1$ and $a_2$ which cannot be solved analytically. These are numerically treated for different parameter sets in chapter 4.2.

To determine the stability of the steady-state solutions the Jacobian corresponding to equations (53) is calculated. The characteristic polynomial is written symbolically as follows:

$$P(\lambda) = \lambda^4 + C_1\lambda^3 + C_2\lambda^2 + C_3\lambda + C_4 \quad (55)$$

According to Routh criterion, stable solutions are achieved for $C_i > 0$, $\Delta_2 = C_1 C_2 - C_3 > 0$ and for $\Delta_3 = C_3\Delta_2 - C_1^2 C_4 > 0$. Hopf bifurcation corresponds to $\lambda = i\omega, \omega \in \mathbb{R}$. That implies $\omega^2 = C_1/C_3$ and $\dfrac{C_3^2}{C_1^2} - \dfrac{C_2 C_3}{C_1} + C_4 = 0 \Rightarrow \Delta_3 = 0$.

Further numerical treatment is presented in section 4.

## 4. Numerical results

### 4.1. Non-resonant case

Equation (50) yields the frequency response of the non-resonant forced system. For different values of $c_{11} = \gamma_{17}\omega_1^2 + 3\alpha_{11}$ we get qualitatively different frequency response graphs[33]. For negative values of $c_{11}$ we get a softening effect (**Fig. 9**(a)), and for positive values of $c_{11}$ we get a hardening effect (**Fig. 9**(b)). The dependence of $c_{11}$ on model coefficient $\kappa$ is shown graphically in **Fig. 10**.

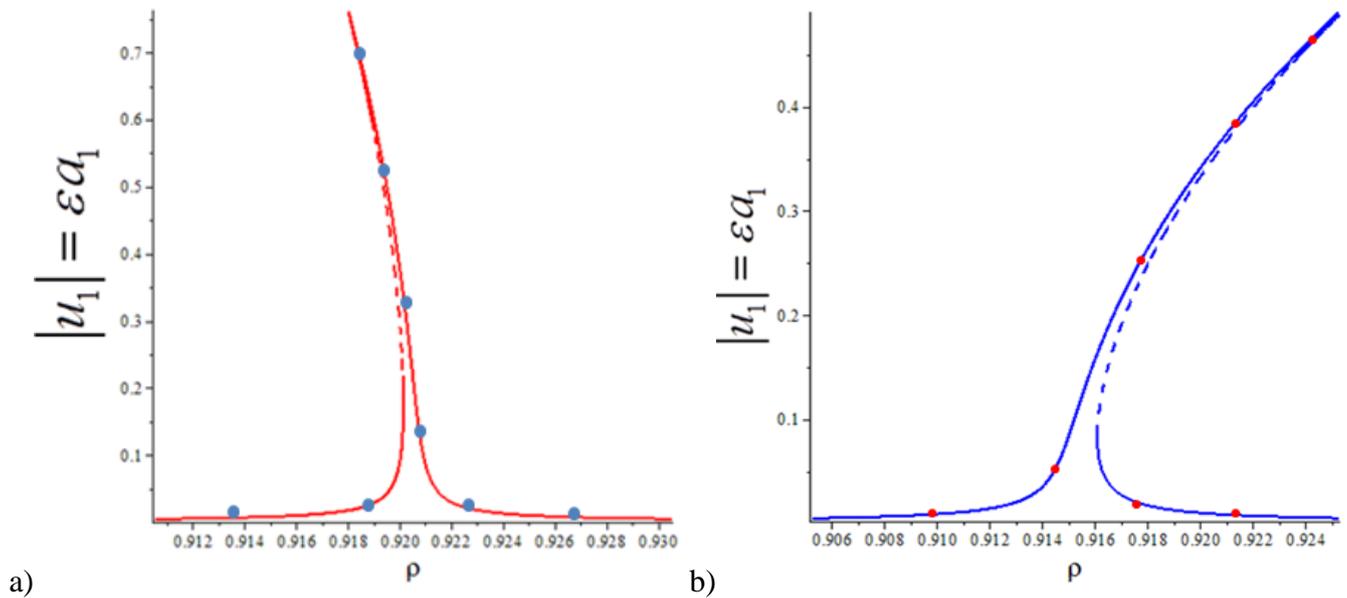

**Fig. 9.** Frequency response graph of $|u_1|$ for $s_1 = 1.5, \mu_1 = 0.5, \theta_0 = \dfrac{\pi}{4}, \gamma = 0.0025, A = 0.0005$ a) $\kappa = 4.7$ and $c_{11} \approx -0.03188$, softening effect; b) $\kappa = 4.5$ and $c_{11} \approx 0.30585$, hardening effect.

In both hardening and softening cases shown in **Fig. 9**, we get a unique stable solution when the excitation frequency $\rho$ is far from the system first natural frequency. However, in the neighborhood of the primary resonance there are three different solutions- two stable solutions (solid) and an unstable solution (dashed).

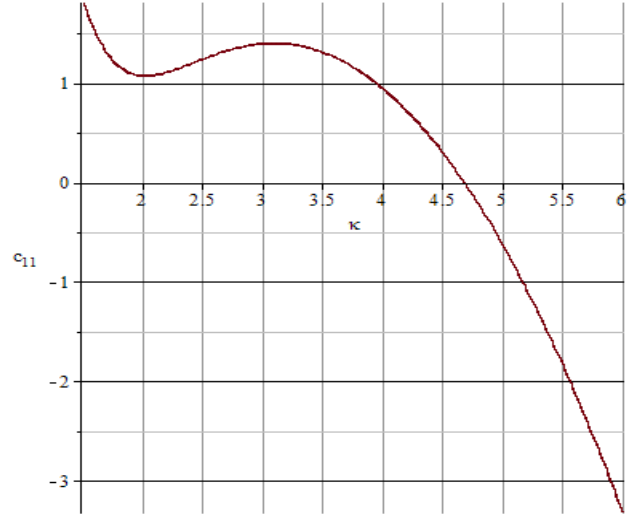

**Fig. 10.** Coefficient $c_{11}$ vs. parameter $\kappa$ for $s_1 = 1.5, \mu_1 = 0.5, \theta_0 = \dfrac{\pi}{4}, \gamma = 0.0025, A = 0.0005$

### 4.2. 3:1 Internal Resonance

Steady-state solutions of equation (53) are calculated for different parameter sets as shown in **Fig. 9**(a) for $\varepsilon^2 \sigma_1 = 0$ and in **Fig. 10** for $\varepsilon^2 \sigma_1 = 0.005$. As one can see in **Fig. 9**, for small values of excitation amplitude, the system has single stable solution. For a critical threshold saddle-node bifurcation takes place, and there exists a region with three solutions - two stable (solid) and one unstable (dashed). Note that the value of the detuning parameter $\sigma_1$ is obtained by changing the value of $\kappa$, which in turn modifies the system natural frequencies according to **Fig. 7**. Parameter $\kappa$ is calculated numerically from equation (51) using equation (28) for chosen values of $\sigma_1$; for $\varepsilon^2 \sigma_1 = 0$ we get $\kappa = 1.808$ (**Fig. 11**), and for $\varepsilon^2 \sigma_1 = 0.005$ we get $\kappa = 1.805$ (**Fig. 12**).

**Fig. 12** demonstrates the bifurcation structure similar to **Fig. 11**. However, the curves are sharper due to lower value of damping and another dynamic region of unstable solution exists between two Hopf bifurcations. In order to determine those Hopf frequency values and determine the branches stability for the case shown in **Fig. 12**, we calculate Routh criterion coefficients $C_1, C_2, C_3, C_4$ and $\Delta_3$ as discussed in section 3.3.2. **Fig. 13** shows Routh criterion coefficient $\Delta_3$ as a function of the normalized external excitation frequency $\rho$. As one can see, for parameter values of $\rho = 0.50925, 0.50998$ coefficient $\Delta_3$ (Routh's criterion) is equal to zero, which means that Hopf bifurcation takes place. Consequently, we expect several effects to occur for these sets of parameters, for both DOFs; quasi-periodic beating signals, close trajectories in state-plane, and Poincare section points forming a closed trajectory.

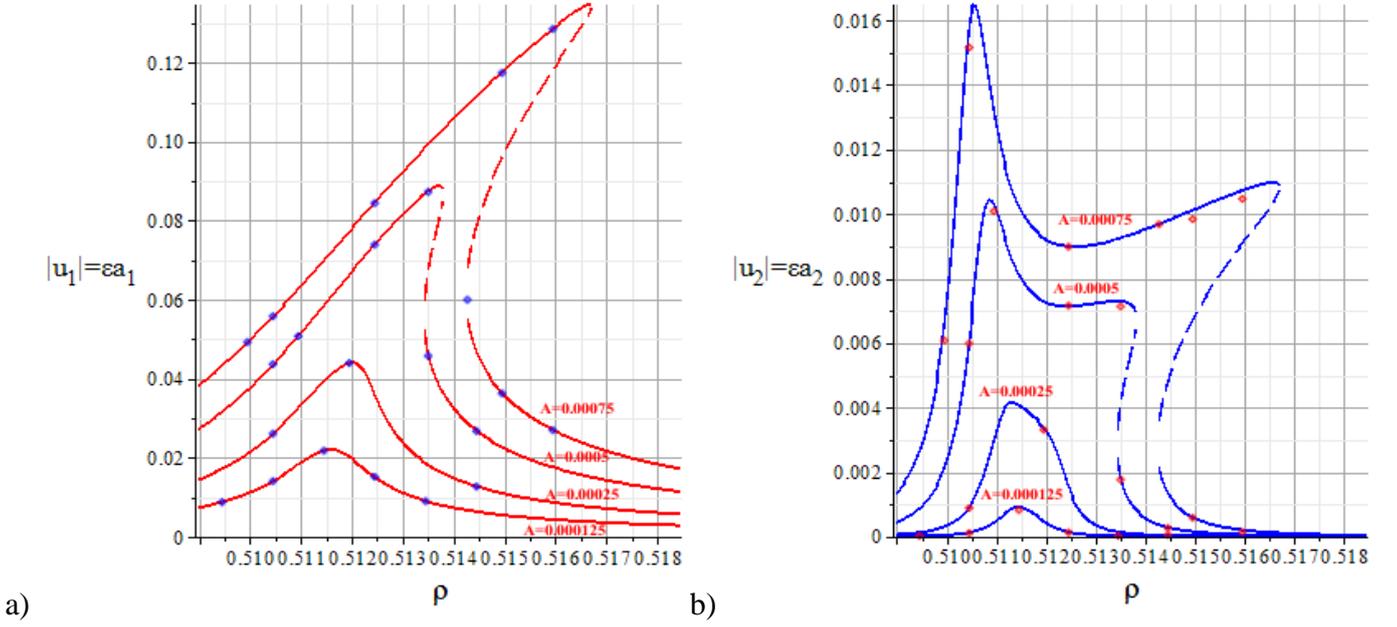

**Fig. 11.** Frequency response curves of $|u_1|$ and $|u_2|$ for $s_1 = 1.5, \theta_0 = \pi/4, \mu_1 = 0.5, \gamma = 0.0025, \varepsilon^2 \sigma_1 = 0$, i.e. $\kappa = 1.81$, under different harmonic excitation amplitudes.

The full equations of motion of system (15) are integrated using Runge-Kutta 4 algorithm to validate the asymptotic results. One can see that in **Fig. 11** and **Fig. 12** there is a good agreement between the asymptotic results and the numerical simulations. For frequency value of $\rho = 0.509615$, where orbital instabilities take place, we observe in **Fig. 14** beats in the time-history signals and closed orbital structure of Poincaré map, which correspond to quasi-periodic motion.

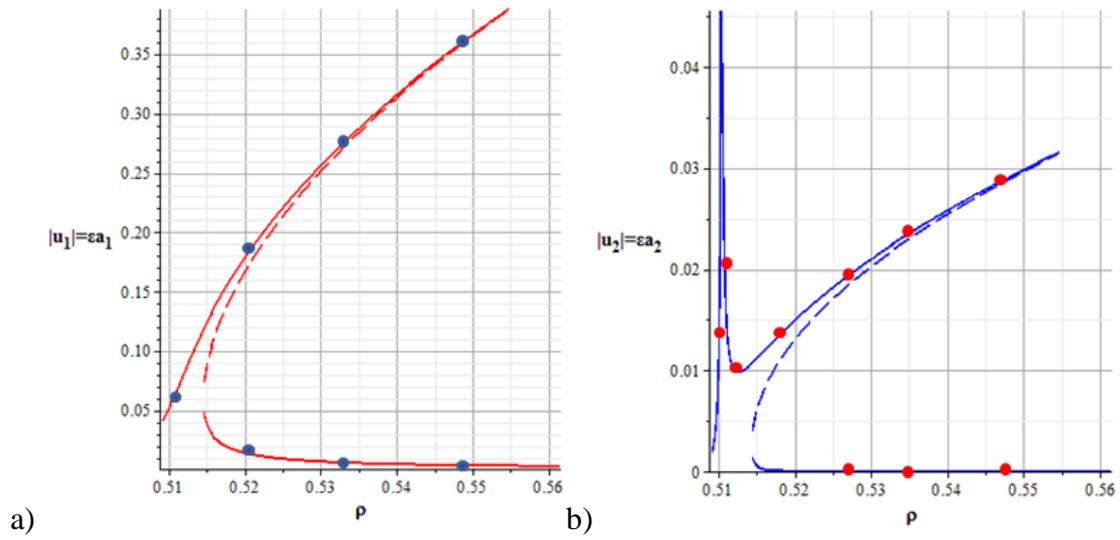

**Fig. 12.** Frequncy response curves of $|u_1|$ and $|u_2|$ for $s_1 = 1.5, \theta_0 = \pi/4, \mu_1 = 0.5, \gamma = 0.001, \varepsilon^2\sigma_1 = 0.005, A = 0.00075$, i.e. $\kappa = 1.8049$

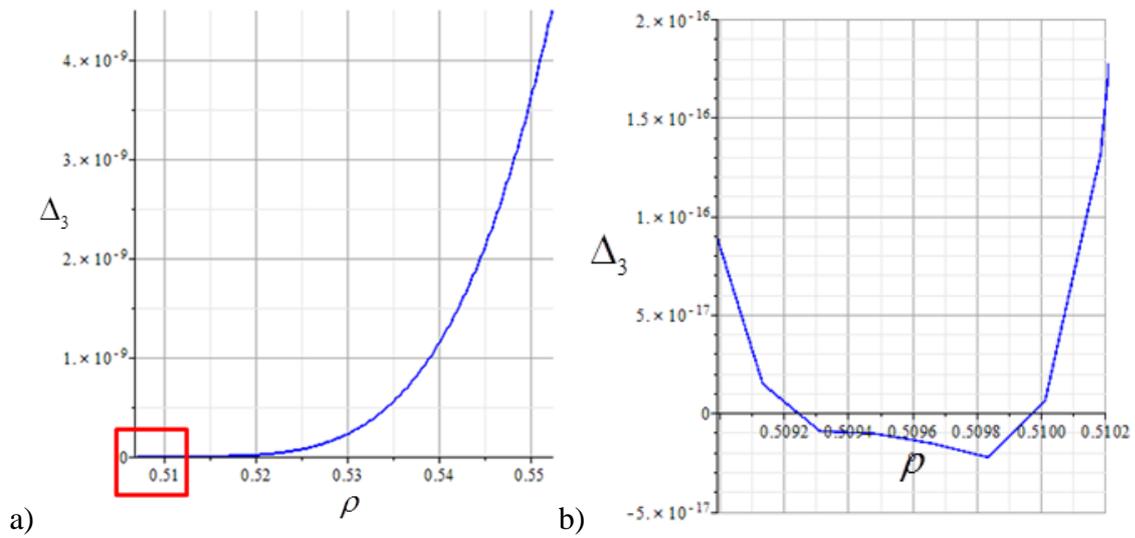

**Fig. 13.** a) Routh coefficient $\Delta_3$ vs. external forcing frequency, b) zoom-in.

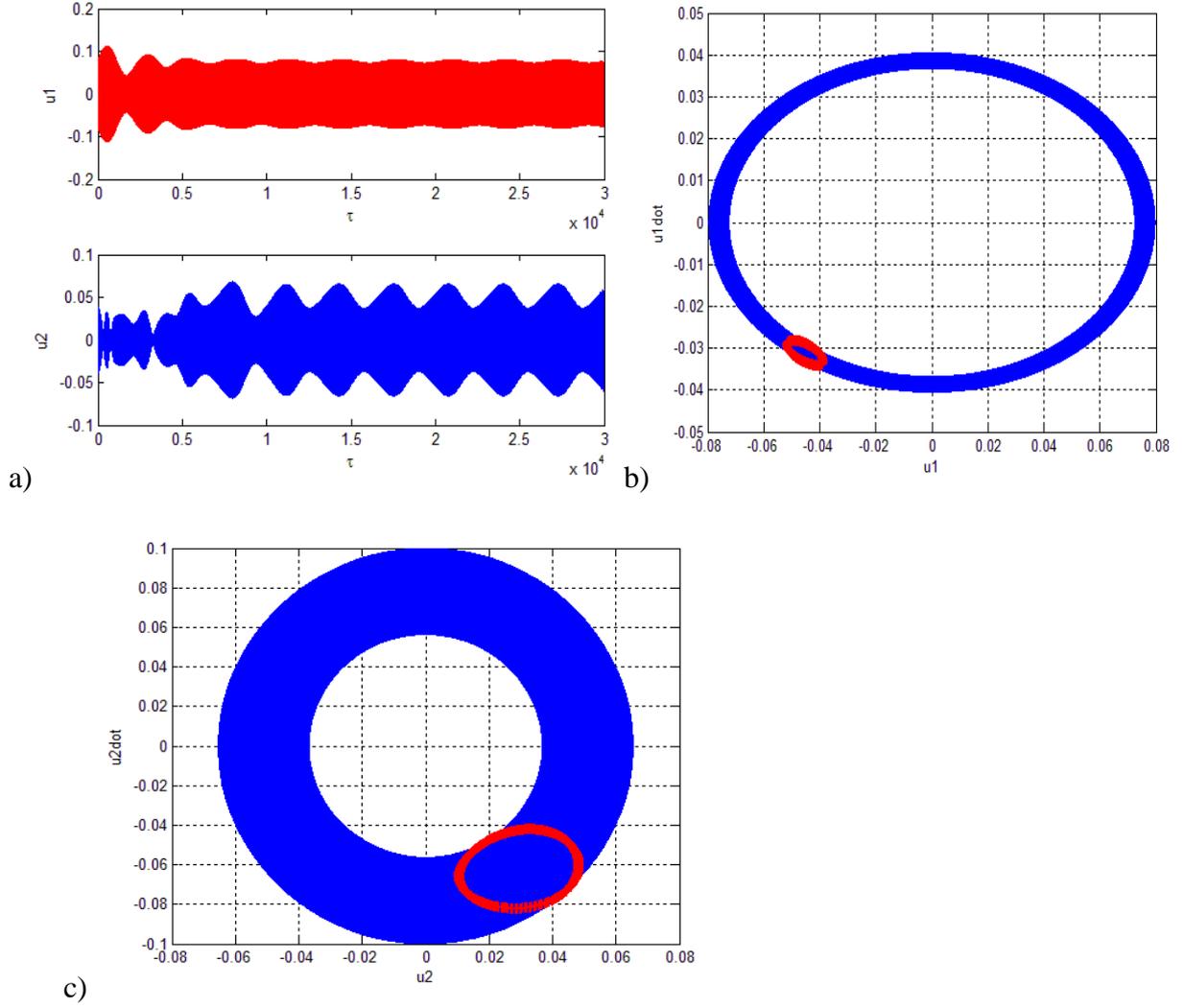

**Fig. 14.** a) Time history for signals $u_1$ and $u_2$. Beating signals take place. b) State-plane and Poincare sections of $u_1$, c) State-plane and Poincare sections of $u_2$. The simulations were performed for parameters: $\mu_1 = 0.5, s_1 = 1.5, \sigma_1 = 0.005, \kappa = 1.804897259, \gamma = 0.001, A = 0.00075, \rho = 0.509615$

To achieve and explore to some extent the regime of impacts, numerical continuations approach is applied through slow variation of the amplitude. This method allows tracing qualitative changes in the system dynamics beyond the range of validity of the asymptotic approximation. We would like to link between small-amplitude regimes revealed asymptotically and high-amplitude response, for which the asymptotic analysis loses its validity. Note that for all farther simulations the natural frequencies are: $\omega_1 = 0.50936$ and $\omega_2 = 1.53308$ and excitation frequency is taken as: $\rho = 0.513$, i.e. in vicinity of 1:1 primary resonance with respect to the lower natural frequency. For small excitation amplitude of A=0.00075 (described in **Fig. 11**), regime of 3:1 internal resonance under primary resonance takes place, as shown in **Fig. 15**. As one can see, for this case, $x_1 \sim O(0.01)$, i.e. the strongly nonlinear effect of impacts is negligible, and the only dominant frequencies are $\omega_1$ and $3\omega_1$, as predicted from the asymptotic analysis.

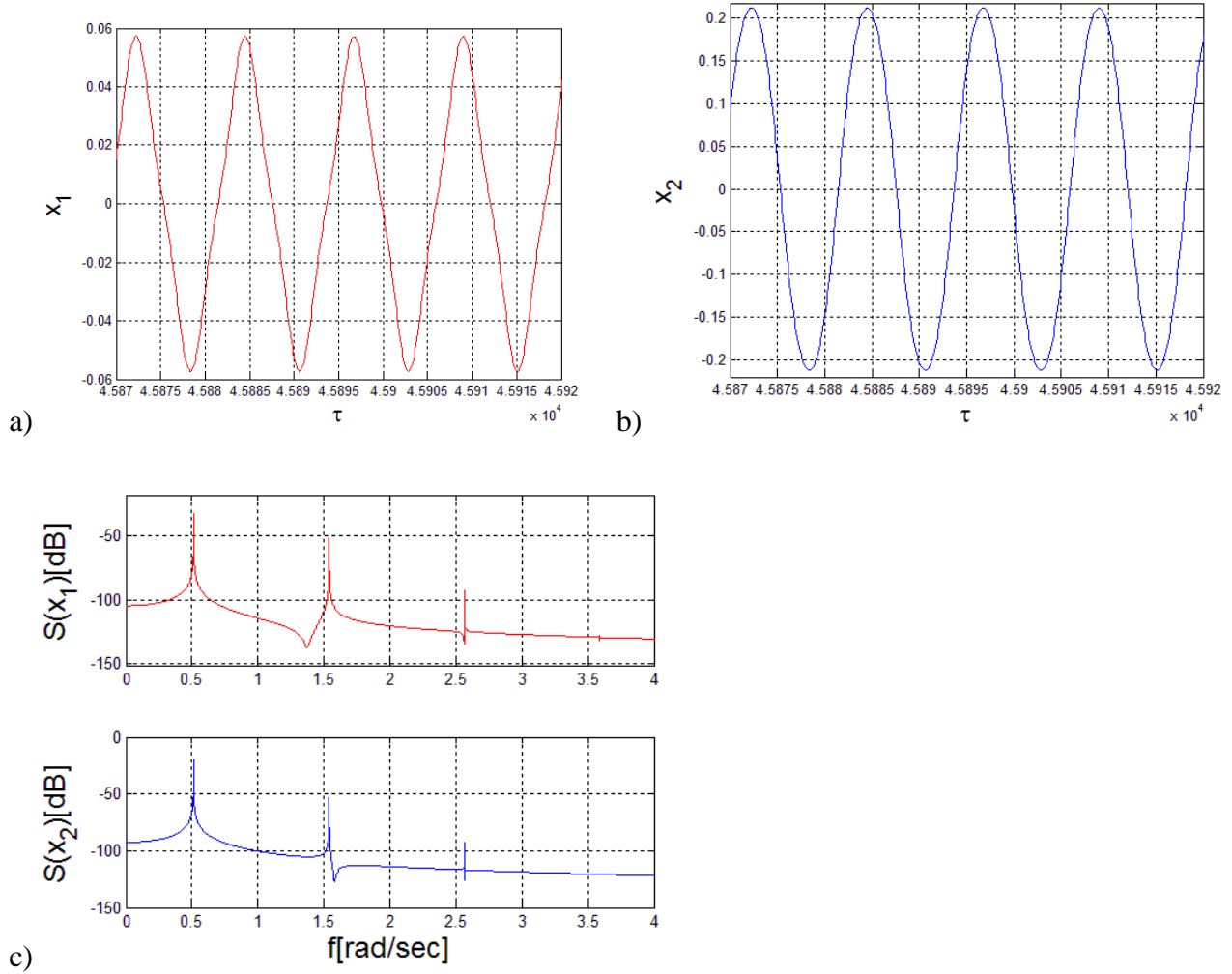

**Fig. 15.** Periodic response given for small excitation amplitude A=0.00075; a)time history of $x_1$, b)time history of $x_2$, c)Fast Fourier transform (FFT) of both signals.

However, for higher amplitude excitation of A=1.0475, $x_1$ is close to unity, and the impact nonlinearity become more significant. Consequently the response contains more frequencies, as shown in **Fig. 16**.

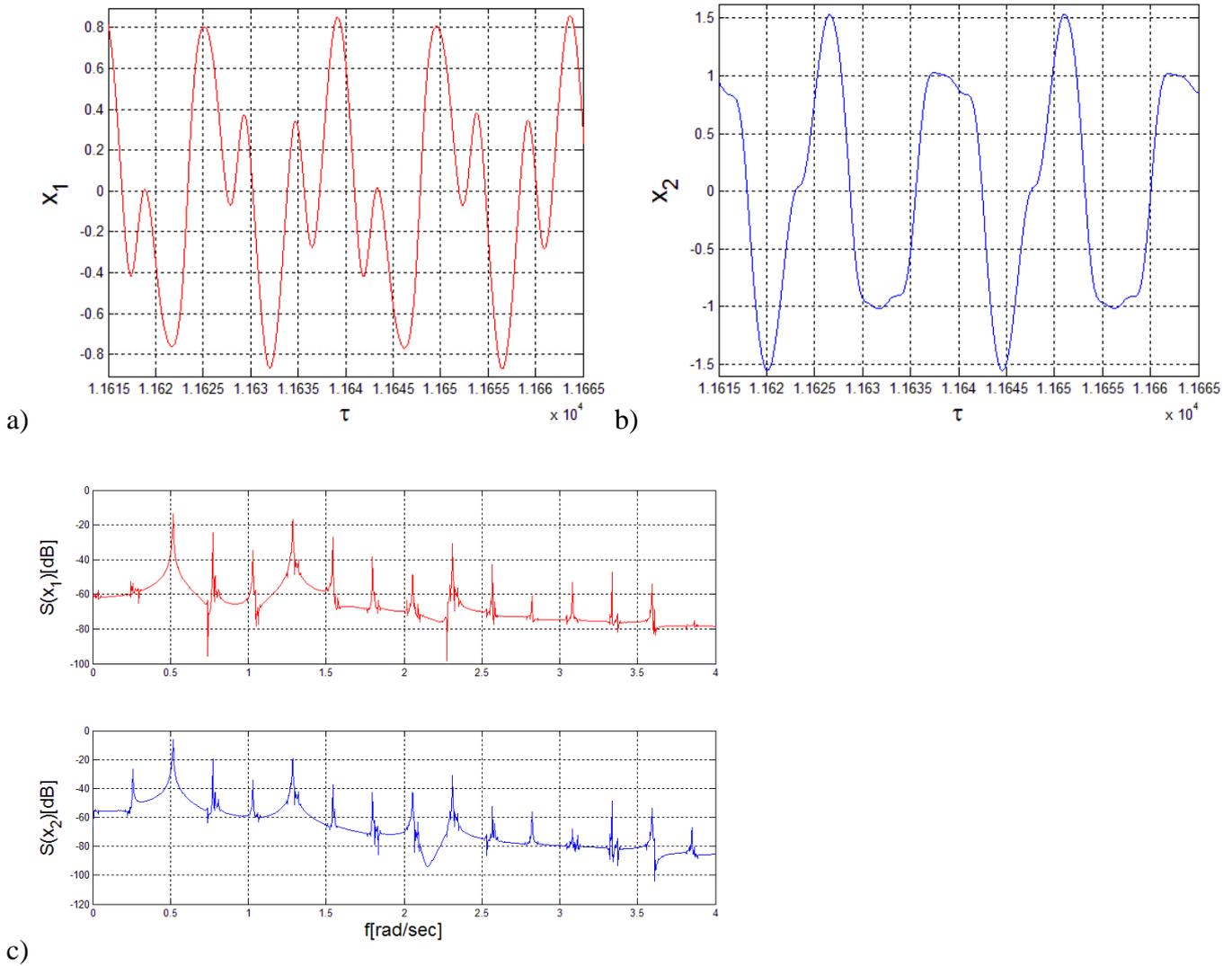

a)

b)

c)

**Fig. 16.** Periodic response given for intermediate excitation amplitude A=1.0475; a)time history of $x_1$, b)time history of $x_2$, c)FFT for $x_1$ time history (up), and for $x_2$ time history(down).

When A=2.0, $x_1$ reaches unity, i.e. hydraulic impacts are present. From **Fig. 17** we learn that this regime is qualitatively identical to the moderate sloshing case presented in **Fig. 16**, albeit with much wider frequency distribution. The latter strongly reminds chaotic motion.

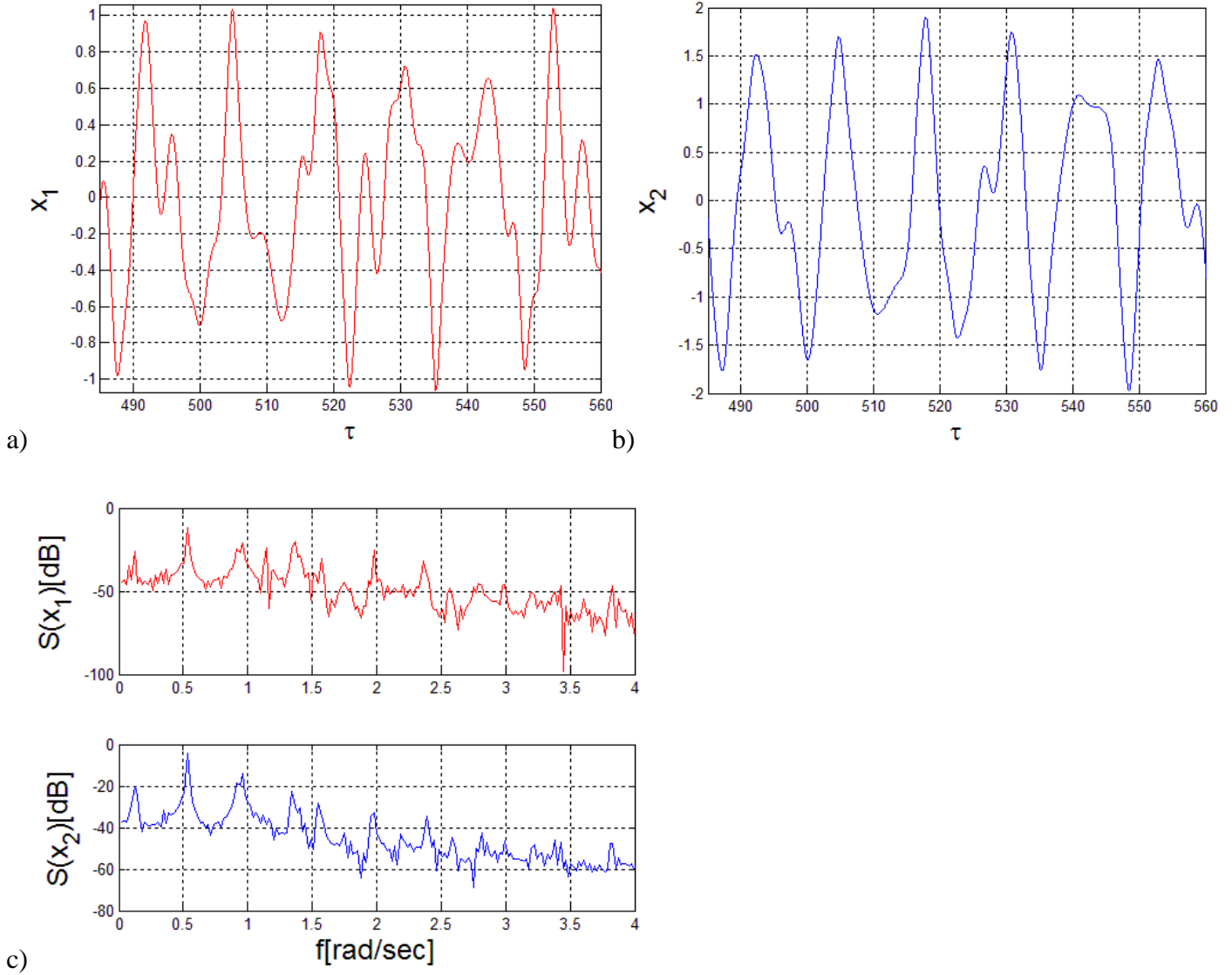

**Fig. 17.** periodic response given for large excitation amplitude A=2.0; a)time history of $x_1$, b)time history of $x_2$, c)FFT for $x_1$ time history (up), and for $x_2$ time history(down).

## 5. Discussion and concluding remarks

We explored the equivalent mechanical model of partially liquid-filled vessel under external horizontal excitation. The system natural frequencies were calculated and conditions for internal resonance were presented. The cases of 1:1 primary resonance without and with 3:1 internal resonance were explored by multiple scales method.

Quasi-periodic motion was revealed for a narrow excitation frequency range. Numerical simulations completely support these findings.

For the same parameter values and under larger excitation amplitude, the system exhibits amplitude response, corresponding to hydraulic impacts on vessel walls. More exactly, for small excitation amplitudes, highly nonlinear terms are negligible and internal resonance 3:1 takes place. When the excitation amplitude becomes larger, the impact terms become more significant. In this case the asymptotic approximation loses its validity, and broadband responses are exhibited in the system dynamics.

The authors are very grateful to Israel Science Foundation (grant 838/13) for financial support.

## Appendix

In this Appendix, some more awkward mathematical expressions from the paper are documented.

$$k_{11} = \frac{1}{1-\mu_1}, \quad k_{12} = \frac{\kappa - s_1}{(1-\mu_1)s_1}, \quad k_{21} = \frac{\mu_1 s_1}{1-\mu_1}, \quad k_{22} = \frac{\kappa - s_1}{1-\mu_1}$$

$$\hat{\gamma}_{11} = \frac{\gamma}{s_1(1-\mu_1)}, \quad \hat{\gamma}_{12} = \frac{\gamma(1+\mu_1)}{2s_1\theta_0(1-\mu_1)}, \quad \hat{\gamma}_{13} = \frac{\theta_0^2 \mu_1}{1-\mu_1}, \quad \hat{\gamma}_{14} = \frac{\gamma\theta_0^2(1+\mu_1)}{2s_1(1-\mu_1)^2}, \quad \hat{\gamma}_{15} = \frac{\theta_0^2}{s_1(1-\mu_1)};$$

$$\hat{\gamma}_{16} = \hat{\gamma}_{13}, \quad \hat{\gamma}_{17} = \hat{\gamma}_{12}, \quad \hat{\gamma}_{18} = \hat{\gamma}_{15}, \quad \hat{\gamma}_{21} = \frac{\gamma}{1-\mu_1}, \quad \hat{\gamma}_{22} = \frac{\gamma\mu_1\theta_0^2}{(1-\mu_1)^2}, \quad \hat{\gamma}_{23} = \frac{\mu_1 s_1 \theta_0^2}{1-\mu_1};$$

$$\hat{\gamma}_{24} = 2\hat{\gamma}_{22}, \quad \hat{\gamma}_{25} = \hat{\gamma}_{13}, \quad \hat{\gamma}_{26} = \hat{\gamma}_{23}, \quad \hat{\gamma}_{27} = \hat{\gamma}_{22}, \quad \hat{\gamma}_{28} = \hat{\gamma}_{25}$$

$$\hat{\alpha}_{11} = \frac{\theta_0^2}{6(1-\mu_1)^2}(5\mu + 1)$$

$$\hat{\alpha}_{12} = \frac{-\theta_0^2}{s_1(1-\mu_1)^2}\left(\frac{1}{2}(1+\mu_1)(\kappa - s_1) - 2s_1\mu_1\right)$$

$$\hat{\alpha}_{13} = \frac{-\theta_0^2}{s_1(1-\mu_1)^2}\left((1+\mu_1)(\kappa - s_1) - s_1\mu_1\right)$$

$$\hat{\alpha}_{14} = \frac{-\theta_0^2}{6s_1(1-\mu_1)^2}\left(3(1+\mu_1)(\kappa - s_1) - s_1(1-\mu_1)\right)$$

$$\hat{\alpha}_{21} = \frac{\theta_0^2 s_1 \mu_1}{3(1-\mu_1)^2}(\mu_1 + 2)$$

$$\hat{\alpha}_{22} = \frac{-\mu_1 \theta_0^2}{(1-\mu_1)^2}\left(s_1(1-\mu_1) - (\kappa - s_1) + 2\mu_1 s_1\right)$$

$$\hat{\alpha}_{23} = \frac{-\mu_1 \theta_0^2}{(1-\mu_1)^2}\left(\frac{1}{2}s_1(1-\mu_1) - 2(\kappa - s_1) + s_1\right)$$

$$\hat{\alpha}_{24} = \frac{-\theta_0^2}{6(1-\mu_1)^2}\left(s_1(1-\mu_1) - 6\mu_1(\kappa - s_1)\right)$$

$$\hat{F}_{12} = \hat{F}_{13} = \frac{\theta_0}{s_1(1-\mu_1)}, \quad \hat{F}_{21} = \frac{1}{\theta_0}, \quad \hat{F}_{22} = \frac{\theta_0 \mu_1}{1-\mu_1}, \quad \hat{F}_{23} = \frac{\theta_0(1+\mu_1)}{2(1-\mu_1)}$$

$$\gamma_{11} = \frac{a}{\omega_2^2 - \omega_1^2}\left(\frac{\omega_2^2 - k_{11}}{k_{12}}\hat{\gamma}_{11} - \hat{\gamma}_{21}\right)$$

$$\gamma_{12} = \frac{b}{\omega_2^2 - \omega_1^2}\left(\frac{\omega_2^2 - k_{11}}{k_{12}}\hat{\gamma}_{11} - \hat{\gamma}_{21}\right)$$

$$\gamma_{13} = \frac{-\left(\frac{\omega_2^2 - k_{11}}{k_{12}}\hat{\gamma}_{12} - \hat{\gamma}_{22}\right)k_{12}^2 b + \left(\frac{\omega_2^2 - k_{11}}{k_{12}}\hat{\gamma}_{14} - \hat{\gamma}_{24}\right)k_{12}ab - \left(\frac{\omega_2^2 - k_{11}}{k_{12}}\hat{\gamma}_{17} - \hat{\gamma}_{27}\right)a^2 b}{\omega_2^2 - \omega_1^2}$$

$$\gamma_{14} = \frac{-\left(\frac{\omega_2^2 - k_{11}}{k_{12}}\hat{\gamma}_{12} - \hat{\gamma}_{22}\right)k_{12}^2 b + \left(\frac{\omega_2^2 - k_{11}}{k_{12}}\hat{\gamma}_{14} - \hat{\gamma}_{24}\right)k_{12}b^2 - \left(\frac{\omega_2^2 - k_{11}}{k_{12}}\hat{\gamma}_{17} - \hat{\gamma}_{27}\right)b^3}{\omega_2^2 - \omega_1^2}$$

$$\gamma_{15} = \frac{-\left(\frac{\omega_2^2 - k_{11}}{k_{12}}\hat{\gamma}_{12} - \hat{\gamma}_{22}\right)k_{12}^2 a + \left(\frac{\omega_2^2 - k_{11}}{k_{12}}\hat{\gamma}_{14} - \hat{\gamma}_{24}\right)k_{12}a^2 - \left(\frac{\omega_2^2 - k_{11}}{k_{12}}\hat{\gamma}_{17} - \hat{\gamma}_{27}\right)a^3}{\omega_2^2 - \omega_1^2}$$

$$\gamma_{16} = \frac{-\left(\frac{\omega_2^2 - k_{11}}{k_{12}}\hat{\gamma}_{12} - \hat{\gamma}_{22}\right)k_{12}^2 a + \left(\frac{\omega_2^2 - k_{11}}{k_{12}}\hat{\gamma}_{14} - \hat{\gamma}_{24}\right)k_{12}ab - \left(\frac{\omega_2^2 - k_{11}}{k_{12}}\hat{\gamma}_{17} - \hat{\gamma}_{27}\right)ab^2}{\omega_2^2 - \omega_1^2}$$

$$\gamma_{17} = \frac{\left(\frac{\omega_2^2 - k_{11}}{k_{12}}\hat{\gamma}_{13} - \hat{\gamma}_{23}\right)k_{12}^3 - \left(\frac{\omega_2^2 - k_{11}}{k_{12}}\hat{\gamma}_{15} - \hat{\gamma}_{25}\right)k_{12}a^2 - \left(\frac{\omega_2^2 - k_{11}}{k_{12}}\hat{\gamma}_{16} - \hat{\gamma}_{26}\right)k_{12}^2 a + \left(\frac{\omega_2^2 - k_{11}}{k_{12}}\hat{\gamma}_{18} - \hat{\gamma}_{28}\right)a^3}{\omega_2^2 - \omega_1^2}$$

$$\gamma_{18} = \frac{2\left(\left(\frac{\omega_2^2 - k_{11}}{k_{12}}\hat{\gamma}_{13} - \hat{\gamma}_{23}\right)k_{12}^3 - \left(\frac{\omega_2^2 - k_{11}}{k_{12}}\hat{\gamma}_{15} - \hat{\gamma}_{25}\right)k_{12}ab - \left(\frac{\omega_2^2 - k_{11}}{k_{12}}\hat{\gamma}_{16} - \hat{\gamma}_{26}\right)k_{12}^2 a + \left(\frac{\omega_2^2 - k_{11}}{k_{12}}\hat{\gamma}_{18} - \hat{\gamma}_{28}\right)a^2 b\right)}{\omega_2^2 - \omega_1^2}$$

$$\gamma_{19} = \frac{\left(\frac{\omega_2^2 - k_{11}}{k_{12}}\hat{\gamma}_{13} - \hat{\gamma}_{23}\right)k_{12}^3 - \left(\frac{\omega_2^2 - k_{11}}{k_{12}}\hat{\gamma}_{15} - \hat{\gamma}_{25}\right)k_{12}b^2 - \left(\frac{\omega_2^2 - k_{11}}{k_{12}}\hat{\gamma}_{16} - \hat{\gamma}_{26}\right)k_{12}^2 a + \left(\frac{\omega_2^2 - k_{11}}{k_{12}}\hat{\gamma}_{18} - \hat{\gamma}_{28}\right)ab^2}{\omega_2^2 - \omega_1^2}$$

$$\gamma_{110} = \frac{2\left(\left(\frac{\omega_2^2 - k_{11}}{k_{12}}\hat{\gamma}_{13} - \hat{\gamma}_{23}\right)k_{12}^3 - \left(\frac{\omega_2^2 - k_{11}}{k_{12}}\hat{\gamma}_{15} - \hat{\gamma}_{25}\right)k_{12}ab - \left(\frac{\omega_2^2 - k_{11}}{k_{12}}\hat{\gamma}_{16} - \hat{\gamma}_{26}\right)k_{12}^2 b + \left(\frac{\omega_2^2 - k_{11}}{k_{12}}\hat{\gamma}_{18} - \hat{\gamma}_{28}\right)ab^2\right)}{\omega_2^2 - \omega_1^2}$$

$$\gamma_{111} = \frac{\left(\frac{\omega_2^2 - k_{11}}{k_{12}}\hat{\gamma}_{13} - \hat{\gamma}_{23}\right)k_{12}^3 - \left(\frac{\omega_2^2 - k_{11}}{k_{12}}\hat{\gamma}_{15} - \hat{\gamma}_{25}\right)k_{12}a^2 - \left(\frac{\omega_2^2 - k_{11}}{k_{12}}\hat{\gamma}_{16} - \hat{\gamma}_{26}\right)k_{12}^2 b + \left(\frac{\omega_2^2 - k_{11}}{k_{12}}\hat{\gamma}_{18} - \hat{\gamma}_{28}\right)a^2 b}{\omega_2^2 - \omega_1^2}$$

$$\gamma_{112} = \frac{-2\left(\frac{\omega_2^2 - k_{11}}{k_{12}}\hat{\gamma}_{12} - \hat{\gamma}_{22}\right)k_{12}^2 a + \left(\frac{\omega_2^2 - k_{11}}{k_{12}}\hat{\gamma}_{14} - \hat{\gamma}_{24}\right)k_{12}a(a+b) - 2\left(\frac{\omega_2^2 - k_{11}}{k_{12}}\hat{\gamma}_{17} - \hat{\gamma}_{27}\right)a^2 b}{\omega_2^2 - \omega_1^2}$$

$$\gamma_{113} = \frac{\left(\frac{\omega_2^2 - k_{11}}{k_{12}}\hat{\gamma}_{13} - \hat{\gamma}_{23}\right)k_{12}^3 - \left(\frac{\omega_2^2 - k_{11}}{k_{12}}\hat{\gamma}_{15} - \hat{\gamma}_{25}\right)k_{12}b^2 - \left(\frac{\omega_2^2 - k_{11}}{k_{12}}\hat{\gamma}_{16} - \hat{\gamma}_{26}\right)k_{12}^2 b + \left(\frac{\omega_2^2 - k_{11}}{k_{12}}\hat{\gamma}_{18} - \hat{\gamma}_{28}\right)b^3}{\omega_2^2 - \omega_1^2}$$

$$\gamma_{114} = \frac{-2\left(\frac{\omega_2^2 - k_{11}}{k_{12}}\hat{\gamma}_{12} - \hat{\gamma}_{22}\right)k_{12}^2 b + \left(\frac{\omega_2^2 - k_{11}}{k_{12}}\hat{\gamma}_{14} - \hat{\gamma}_{24}\right)k_{12}b(a+b) - 2\left(\frac{\omega_2^2 - k_{11}}{k_{12}}\hat{\gamma}_{17} - \hat{\gamma}_{27}\right)ab^2}{\omega_2^2 - \omega_1^2}$$

where $a = \omega_1^2 - k_{11}$ and $b = \omega_2^2 - k_{11}$.

$$\gamma_{21} = \frac{-a}{\omega_2^2 - \omega_1^2}\left(\frac{\omega_1^2 - k_{11}}{k_{12}}\hat{\gamma}_{11} - \hat{\gamma}_{21}\right)$$

$$\gamma_{22} = \frac{-b}{\omega_2^2 - \omega_1^2}\left(\frac{\omega_1^2 - k_{11}}{k_{12}}\hat{\gamma}_{11} - \hat{\gamma}_{21}\right)$$

$$\gamma_{23} = -\frac{-k_{12}^2 b\left(\frac{\omega_1^2 - k_{11}}{k_{12}}\hat{\gamma}_{12} - \hat{\gamma}_{22}\right) + k_{12}ab\left(\frac{\omega_1^2 - k_{11}}{k_{12}}\hat{\gamma}_{14} - \hat{\gamma}_{24}\right) - a^2 b\left(\frac{\omega_1^2 - k_{11}}{k_{12}}\hat{\gamma}_{17} - \hat{\gamma}_{27}\right)}{\omega_2^2 - \omega_1^2}$$

$$\gamma_{24} = -\frac{-k_{12}^2 b\left(\frac{\omega_1^2 - k_{11}}{k_{12}}\hat{\gamma}_{12} - \hat{\gamma}_{22}\right) + k_{12}b^2\left(\frac{\omega_1^2 - k_{11}}{k_{12}}\hat{\gamma}_{14} - \hat{\gamma}_{24}\right) - b^3\left(\frac{\omega_1^2 - k_{11}}{k_{12}}\hat{\gamma}_{17} - \hat{\gamma}_{27}\right)}{\omega_2^2 - \omega_1^2}$$

$$\gamma_{25} = -\frac{-k_{12}^2 a\left(\frac{\omega_1^2 - k_{11}}{k_{12}}\hat{\gamma}_{12} - \hat{\gamma}_{22}\right) + k_{12}a^2\left(\frac{\omega_1^2 - k_{11}}{k_{12}}\hat{\gamma}_{14} - \hat{\gamma}_{24}\right) - a^3\left(\frac{\omega_1^2 - k_{11}}{k_{12}}\hat{\gamma}_{17} - \hat{\gamma}_{27}\right)}{\omega_2^2 - \omega_1^2}$$

$$\gamma_{26} = -\frac{-k_{12}^2 a\left(\frac{\omega_1^2 - k_{11}}{k_{12}}\hat{\gamma}_{12} - \hat{\gamma}_{22}\right) + k_{12}ab\left(\frac{\omega_1^2 - k_{11}}{k_{12}}\hat{\gamma}_{14} - \hat{\gamma}_{24}\right) - ab^2\left(\frac{\omega_1^2 - k_{11}}{k_{12}}\hat{\gamma}_{17} - \hat{\gamma}_{27}\right)}{\omega_2^2 - \omega_1^2}$$

$$\gamma_{27} = -\frac{k_{12}^3\left(\frac{\omega_1^2 - k_{11}}{k_{12}}\hat{\gamma}_{13} - \hat{\gamma}_{23}\right) - k_{12}a^2\left(\frac{\omega_1^2 - k_{11}}{k_{12}}\hat{\gamma}_{15} - \hat{\gamma}_{25}\right) - k_{12}^2 a\left(\frac{\omega_1^2 - k_{11}}{k_{12}}\hat{\gamma}_{16} - \hat{\gamma}_{26}\right) + a^3\left(\frac{\omega_1^2 - k_{11}}{k_{12}}\hat{\gamma}_{18} - \hat{\gamma}_{28}\right)}{\omega_2^2 - \omega_1^2}$$

$$\gamma_{28} = -\frac{2\left(k_{12}^3\left(\frac{\omega_1^2 - k_{11}}{k_{12}}\hat{\gamma}_{13} - \hat{\gamma}_{23}\right) - k_{12}ab\left(\frac{\omega_1^2 - k_{11}}{k_{12}}\hat{\gamma}_{15} - \hat{\gamma}_{25}\right) - k_{12}^2 a\left(\frac{\omega_1^2 - k_{11}}{k_{12}}\hat{\gamma}_{16} - \hat{\gamma}_{26}\right) + a^2 b\left(\frac{\omega_1^2 - k_{11}}{k_{12}}\hat{\gamma}_{18} - \hat{\gamma}_{28}\right)\right)}{\omega_2^2 - \omega_1^2}$$

$$\gamma_{29} = -\frac{k_{12}^3\left(\frac{\omega_1^2 - k_{11}}{k_{12}}\hat{\gamma}_{13} - \hat{\gamma}_{23}\right) - k_{12}b^2\left(\frac{\omega_1^2 - k_{11}}{k_{12}}\hat{\gamma}_{15} - \hat{\gamma}_{25}\right) - k_{12}^2 a\left(\frac{\omega_1^2 - k_{11}}{k_{12}}\hat{\gamma}_{16} - \hat{\gamma}_{26}\right) + ab^2\left(\frac{\omega_1^2 - k_{11}}{k_{12}}\hat{\gamma}_{18} - \hat{\gamma}_{28}\right)}{\omega_2^2 - \omega_1^2}$$

$$\gamma_{210} = -\frac{2\left(k_{12}^3\left(\frac{\omega_1^2 - k_{11}}{k_{12}}\hat{\gamma}_{13} - \hat{\gamma}_{23}\right) - k_{12}ab\left(\frac{\omega_1^2 - k_{11}}{k_{12}}\hat{\gamma}_{15} - \hat{\gamma}_{25}\right) - k_{12}^2 b\left(\frac{\omega_1^2 - k_{11}}{k_{12}}\hat{\gamma}_{16} - \hat{\gamma}_{26}\right) + ab^2\left(\frac{\omega_1^2 - k_{11}}{k_{12}}\hat{\gamma}_{18} - \hat{\gamma}_{28}\right)\right)}{\omega_2^2 - \omega_1^2}$$

$$\gamma_{211} = -\frac{k_{12}^3\left(\frac{\omega_1^2 - k_{11}}{k_{12}}\hat{\gamma}_{13} - \hat{\gamma}_{23}\right) - k_{12}a^2\left(\frac{\omega_1^2 - k_{11}}{k_{12}}\hat{\gamma}_{15} - \hat{\gamma}_{25}\right) - k_{12}^2 b\left(\frac{\omega_1^2 - k_{11}}{k_{12}}\hat{\gamma}_{16} - \hat{\gamma}_{26}\right) + a^2 b\left(\frac{\omega_1^2 - k_{11}}{k_{12}}\hat{\gamma}_{18} - \hat{\gamma}_{28}\right)}{\omega_2^2 - \omega_1^2}$$

$$\gamma_{212} = -\frac{-2k_{12}^2 a\left(\frac{\omega_1^2 - k_{11}}{k_{12}}\hat{\gamma}_{12} - \hat{\gamma}_{22}\right) + k_{12}a(a+b)\left(\frac{\omega_1^2 - k_{11}}{k_{12}}\hat{\gamma}_{14} - \hat{\gamma}_{24}\right) - 2a^2 b\left(\frac{\omega_1^2 - k_{11}}{k_{12}}\hat{\gamma}_{17} - \hat{\gamma}_{27}\right)}{\omega_2^2 - \omega_1^2}$$

$$\gamma_{213} = -\frac{k_{12}^3\left(\frac{\omega_1^2 - k_{11}}{k_{12}}\hat{\gamma}_{13} - \hat{\gamma}_{23}\right) - k_{12}b^2\left(\frac{\omega_1^2 - k_{11}}{k_{12}}\hat{\gamma}_{15} - \hat{\gamma}_{25}\right) - k_{12}^2 b\left(\frac{\omega_1^2 - k_{11}}{k_{12}}\hat{\gamma}_{16} - \hat{\gamma}_{26}\right) + b^3\left(\frac{\omega_1^2 - k_{11}}{k_{12}}\hat{\gamma}_{18} - \hat{\gamma}_{28}\right)}{\omega_2^2 - \omega_1^2}$$

$$\gamma_{214} = -\frac{-2k_{12}^2 b\left(\frac{\omega_1^2 - k_{11}}{k_{12}}\hat{\gamma}_{12} - \hat{\gamma}_{22}\right) + k_{12}b(a+b)\left(\frac{\omega_1^2 - k_{11}}{k_{12}}\hat{\gamma}_{14} - \hat{\gamma}_{24}\right) - 2ab^2\left(\frac{\omega_1^2 - k_{11}}{k_{12}}\hat{\gamma}_{17} - \hat{\gamma}_{27}\right)}{\omega_2^2 - \omega_1^2}$$

$$\alpha_{11} = \frac{-\left(\frac{\omega_2^2 - k_{11}}{k_{12}}\hat{\alpha}_{11} - \hat{\alpha}_{21}\right)k_{12}^3 + a\left(\frac{\omega_2^2 - k_{11}}{k_{12}}\hat{\alpha}_{12} + \hat{\alpha}_{22}\right)k_{12}^2 - \left(\frac{\omega_2^2 - k_{11}}{k_{12}}\hat{\alpha}_{13} + \hat{\alpha}_{23}\right)k_{12}a^2 + \left(\frac{\omega_2^2 - k_{11}}{k_{12}}\hat{\alpha}_{14} + \hat{\alpha}_{24}\right)a^3}{\omega_2^2 - \omega_1^2}$$

$$\alpha_{12} = \frac{-\left(\frac{\omega_2^2 - k_{11}}{k_{12}}\hat{\alpha}_{11} - \hat{\alpha}_{21}\right)k_{12}^3 + \left(\frac{\omega_2^2 - k_{11}}{k_{12}}\hat{\alpha}_{12} + \hat{\alpha}_{22}\right)k_{12}^2 b - \left(\frac{\omega_2^2 - k_{11}}{k_{12}}\hat{\alpha}_{13} + \hat{\alpha}_{23}\right)k_{12}b^2 + \left(\frac{\omega_2^2 - k_{11}}{k_{12}}\hat{\alpha}_{14} + \hat{\alpha}_{24}\right)b^3}{\omega_2^2 - \omega_1^2}$$

$$\alpha_{13} = \frac{-3\left(\frac{\omega_2^2 - k_{11}}{k_{12}}\hat{\alpha}_{11} - \hat{\alpha}_{21}\right)k_{12}^3 + \left(\frac{\omega_2^2 - k_{11}}{k_{12}}\hat{\alpha}_{12} + \hat{\alpha}_{22}\right)k_{12}^2(b+2a) - \left(\frac{\omega_2^2 - k_{11}}{k_{12}}\hat{\alpha}_{13} + \hat{\alpha}_{23}\right)k_{12}\left(2ab + a^2\right) + 3\left(\frac{\omega_2^2 - k_{11}}{k_{12}}\hat{\alpha}_{14} + \hat{\alpha}_{24}\right)a^2 b}{\omega_2^2 - \omega_1^2}$$

$$\alpha_{14} = \frac{-3\left(\frac{\omega_2^2 - k_{11}}{k_{12}}\hat{\alpha}_{11} - \hat{\alpha}_{21}\right)k_{12}^3 + \left(\frac{\omega_2^2 - k_{11}}{k_{12}}\hat{\alpha}_{12} + \hat{\alpha}_{22}\right)k_{12}^2(a+2b) - \left(\frac{\omega_2^2 - k_{11}}{k_{12}}\hat{\alpha}_{13} + \hat{\alpha}_{23}\right)k_{12}\left(2ab + b^2\right) + 3\left(\frac{\omega_2^2 - k_{11}}{k_{12}}\hat{\alpha}_{14} + \hat{\alpha}_{24}\right)ab^2}{\omega_2^2 - \omega_1^2}$$

$$\alpha_{21} = -\frac{-k_{12}^3\left(\frac{\omega_1^2 - k_{11}}{k_{12}}\hat{\alpha}_{11} - \hat{\alpha}_{21}\right) + a\left(\frac{\omega_1^2 - k_{11}}{k_{12}}\hat{\alpha}_{12} + \hat{\alpha}_{22}\right)k_{12}^2 - \left(\frac{\omega_1^2 - k_{11}}{k_{12}}\hat{\alpha}_{13} + \hat{\alpha}_{23}\right)k_{12}a^2 + \left(\frac{\omega_1^2 - k_{11}}{k_{12}}\hat{\alpha}_{14} + \hat{\alpha}_{24}\right)a^3}{\omega_2^2 - \omega_1^2}$$

$$\alpha_{22} = -\frac{-k_{12}^3\left(\frac{\omega_1^2 - k_{11}}{k_{12}}\hat{\alpha}_{11} - \hat{\alpha}_{21}\right) + k_{12}^2 b\left(\frac{\omega_1^2 - k_{11}}{k_{12}}\hat{\alpha}_{12} + \hat{\alpha}_{22}\right) - k_{12}b^2\left(\frac{\omega_1^2 - k_{11}}{k_{12}}\hat{\alpha}_{13} + \hat{\alpha}_{23}\right) + \left(\frac{\omega_1^2 - k_{11}}{k_{12}}\hat{\alpha}_{14} + \hat{\alpha}_{24}\right)b^3}{\omega_2^2 - \omega_1^2}$$

$$\alpha_{23} = -\frac{-3k_{12}^3\left(\frac{\omega_1^2 - k_{11}}{k_{12}}\hat{\alpha}_{11} - \hat{\alpha}_{21}\right) + k_{12}^2(b+2a)\left(\frac{\omega_1^2 - k_{11}}{k_{12}}\hat{\alpha}_{12} + \hat{\alpha}_{22}\right) - k_{12}\left(2ab + a^2\right)\left(\frac{\omega_1^2 - k_{11}}{k_{12}}\hat{\alpha}_{13} + \hat{\alpha}_{23}\right) + 3\left(\frac{\omega_1^2 - k_{11}}{k_{12}}\hat{\alpha}_{14} + \hat{\alpha}_{24}\right)a^2 b}{\omega_2^2 - \omega_1^2}$$

$$\alpha_{24} = -\frac{-3k_{12}^3\left(\frac{\omega_1^2 - k_{11}}{k_{12}}\hat{\alpha}_{11} - \hat{\alpha}_{21}\right) + k_{12}^2(a+2b)\left(\frac{\omega_1^2 - k_{11}}{k_{12}}\hat{\alpha}_{12} + \hat{\alpha}_{22}\right) - k_{12}\left(2ab + b^2\right)\left(\frac{\omega_1^2 - k_{11}}{k_{12}}\hat{\alpha}_{13} + \hat{\alpha}_{23}\right) + 3\left(\frac{\omega_1^2 - k_{11}}{k_{12}}\hat{\alpha}_{14} + \hat{\alpha}_{24}\right)ab^2}{\omega_2^2 - \omega_1^2}$$

$$F_{11} = \frac{\hat{F}_{21}}{\omega_2^2 - \omega_1^2}$$

$$F_{12} = \frac{1}{\omega_2^2 - \omega_1^2}\left(b(a+b)\hat{F}_{12} - \frac{2ab^2}{k_{12}}\hat{F}_{13} - k_{12}(a+b)\hat{F}_{22} + 2ab\hat{F}_{23}\right)$$

$$F_{13} = \frac{1}{\omega_2^2 - \omega_1^2}\left(b^2\hat{F}_{12} - \frac{b^3}{k_{12}}\hat{F}_{13} - k_{12}b\hat{F}_{22} + b^2\hat{F}_{23}\right)$$

$$F_{14} = \frac{1}{\omega_2^2 - \omega_1^2}\left(ab\hat{F}_{12} - \frac{a^2 b}{k_{12}}\hat{F}_{13} - k_{12}a\hat{F}_{22} + a^2\hat{F}_{23}\right)$$

$$F_{22} = \frac{1}{\omega_2^2 - \omega_1^2}\left(-a(a+b)\hat{F}_{12} + \frac{2a^2 b}{k_{12}}\hat{F}_{13} + k_{12}(a+b)\hat{F}_{22} - 2ab\hat{F}_{23}\right)$$

$$F_{23} = \frac{1}{\omega_2^2 - \omega_1^2}\left(-ab\hat{F}_{12} + \frac{ab^2}{k_{12}}\hat{F}_{13} + k_{12}b\hat{F}_{22} - b^2\hat{F}_{23}\right)$$

$$F_{24} = \frac{1}{\omega_2^2 - \omega_1^2}\left(-a^2\hat{F}_{12} + \frac{a^3}{k_{12}}\hat{F}_{13} + k_{12}a\hat{F}_{22} - a^2\hat{F}_{23}\right)$$

$$c_{11} = \gamma_{17}\omega_1^2 + 3\alpha_{11}$$
$$c_{12} = \alpha_{14} + \gamma_{19}\omega_2^2$$
$$c_{13} = -\gamma_{18}\omega_1\omega_2 + \gamma_{111}\omega_1^2 - \alpha_{13}$$
$$c_{21} = 3\alpha_{22} + \gamma_{213}\omega_2^2$$
$$c_{22} = \gamma_{211}\omega_1^2 + \alpha_{23}$$
$$c_{23} = \gamma_{27}\omega_1^2 - \alpha_{21}$$